\documentclass{emulateapj}

\newcommand{\qsoname}  {SDSS J104257.58+074850.5}
\newcommand{\galaxyname} {GQ1042+0747}
\newcommand{\HI}{\ion{H}{1}~}
\newcommand{\kms}{km~s$^{-1}$ }
\newcommand{\Lya}{Lyman~$\alpha$~}

\slugcomment{ }

\shorttitle{Detection of Low-z 21cm \HI Absorber}
\shortauthors{Borthakur et al.}

\begin{document}

\title{ Using 21cm Absorption in Small Impact Parameter Galaxy-QSO Pairs to Probe Low-Redshift Damped and Sub-Damped Lyman $\alpha$ Systems\altaffilmark{1}}

\altaffiltext{1}{Based on observations with (1) the telescopes of the
National Radio Astronomy Observatory, a facility of the National
Science Foundation operated under cooperative agreement by Associated
Universities, Inc., (2) the SOAR Telescope, a joint project of
Conselho Nacional de Pesquisas Cientificas e Tecnoligicas CNPq-Brazil,
The University of North Carolina at Chapel Hill, Michigan State
University, and the National Optical Astronomy Observatory, and (3)
the Apache Point Observatory 3.5m telescope, which is owned and
operated by the Astrophysical Research Consortium.}

\author{Sanchayeeta Borthakur, Todd M. Tripp, Min S. Yun}
\affil{Department of Astronomy, University of Massachusetts, Amherst, MA 01003, USA}
\email{sanch@astro.umass.edu}

\author{Emmanuel Momjian}
\affil{National Radio Astronomy Observatory, 1003 Lopezville Road, Socorro, NM 87801, USA}

\author{Joseph D. Meiring\altaffilmark{2}}
\affil{Department of Physics and Astronomy, University of Louisville, Louisville, KY 40292}
\altaffiltext{2}{Visiting astronomer, Cerro Tololo Inter-American
Observatory and National Optical Astronomy Observatory, which are
operated by the Association of Universities for Research in Astronomy,
under contract with the National Science Foundation.}

\author{David V. Bowen }
\affil{Princeton University Observatory, Peyton Hall, Ivy Lane, Princeton NJ 08544}

\author{Donald G. York }
\affil{Department of Astronomy and Astrophysics, University of Chicago, Chicago, IL 60637, USA; Enrico Fermi Institute, University of Chicago, Chicago, IL 60637, USA}

\begin{abstract}
To search for low-redshift damped Lyman $\alpha$ (DLA) and sub-DLA
quasar absorbers, we have conducted a 21cm absorption survey of
radio-loud quasars at small impact parameters to foreground galaxies
selected from the Sloan Digital Sky Survey (SDSS). Here we present the
first results from this survey based on observations of \qsoname ~
($z_{\rm QSO}$ = 2.66521), a quasar at an angular separation from a
foreground galaxy ($z_{\rm gal}$ = 0.03321) of 2.5$^{\prime\prime}$
(1.7 kpc in projection).  The foreground galaxy is a low-luminosity
spiral with on-going star formation (0.004~M$_{\odot}$~yr$^{-1}$~kpc$^{-2}$) 
and a metallicity of $-0.27 \pm
0.05$ dex.  We detect 21cm absorption from the galaxy with the Green
Bank Telescope (GBT), the Very Large Array (VLA), and the Very Long
Baseline Array (VLBA). The absorption appears to be quiescent disk gas
corotating with the galaxy and we do not find any evidence for outflowing cold neutral gas. The width of the main absorption line
indicates that the gas is cold, $T_{k} < 283$ K, and the \ion{H}{1}
column is surprisingly low given the impact parameter of 1.7 kpc; we
find that $N$(\ion{H}{1})~$\leq~9.6~\times~10^{19}$~cm$^{-2}$ (GBT)
and $N$(\ion{H}{1})~$\leq~1.5~\times~10^{20}$~cm$^{-2}$ (VLBA). VLBA marginally resolves the continuum source and the absorber, and a lower limit of 27.1 $\times$ 13.9~pc  is derived for the size of the absorbing cloud. In turn, this indicates a low density for a
cold cloud, $n$(\ion{H}{1})~$<$~3.5~cm$^{-3}$.  We hypothesize that this
galaxy, which is relatively isolated, is becoming depleted in
\ion{H}{1} because it is converting its interstellar matter into stars
without a replenishing source of gas, and we suggest future
observations to probe this and similar galaxies.

\end{abstract}


\section{Introduction\label{intro}}

Damped Lyman-$\alpha$ absorbers [DLAs; defined by \citet {wolfe86} as $N$(\ion{H}{1}) $> 2 \times 10^{20}$ cm$^{-2}$] and sub-Damped Lyman-$\alpha$ absorbers [sub-DLAs ; defined by \citet{peroux01} as  1.6$~ \times ~10^{17}$ cm$^{-2}~ <~  N$(\ion{H}{1}) $< 2 \times 10^{20}$ cm$^{-2}$]  are important probes of galaxy evolution for several reasons:

First, the high \ion{H}{1} column density in DLAs shields the
absorbers from photoionization by ultraviolet light, which allows the
gas to cool sufficiently so that these systems can be major sites of
star formation. Moreover, many studies have shown that DLAs are the
dominant reservoirs of {\it neutral} gas throughout most of the
history of the universe \citep[e.g.,][]{proc05}, and it has been
suggested that DLAs are the progenitors of modern-day disk galaxies
\citep{wolfe86}.  The most recent DLA studies employing large
statistical samples \citep{pw09, noter09a} indicate that the
cosmological mass density of neutral gas in DLAs ($\Omega^{\rm DLA}
_{g}$) decreases with redshift from $z = 4$ down to $z = 2.2$, but
there is controversy about how $\Omega^{\rm DLA} _{g}$ evolves at $z
\ll 2$.  The decrease in $\Omega^{\rm DLA} _{g}$ could reflect the
consumption of gas by conversion to stars, but \citet{pw09} argue that
a more complicated picture involving both inflows and outflows is
suggested by the data.

Second, the self-shielding of the DLAs facilitates accurate
metallicity measurements by removing the substantial uncertainties
caused by ionization corrections in lower-$N$(\ion{H}{1}) absorption
systems.  In principle, DLAs do not suffer from any luminosity bias
either; DLA metallicity measurements can be carried out equally well
at any redshift as long as a higher$-z$ background QSO can be found.
Thus, DLAs are powerful tools for tracing the chemical enrichment
history of galaxies and their progenitors.  The higher \ion{H}{1}
column also enables detection of weaker lines and exotic
species \citep[e.g.,][]{proc03,junkk04,york06a,pettini08,ellison08}, and
detailed signatures of various nucleosynthetic processes can be
investigated in DLAs as well as those from dust extinction.

Third, by using DLAs to select high-redshift gas-rich galaxies, the
kinematics of high$-z$ galactic gas flows can be investigated in
detail, and with good statistical samples, based on the kinematics of
sensitive metal absorption lines \citep{pw97,proc08}.  Several groups
have used simulations and the observed DLA kinematics to argue that
gravity-driven infall is not sufficient to produce the observed
velocity spreads of metal lines in DLAs, and some additional process
(or processes) such as galactic winds appears to be required
\citep{razo08,pontz08}.

However, in most cases little or no information is available regarding
the {\it origin and environment} in which damped \Lya absorption arises, and this
has hampered the exploitation of DLAs for probing galaxy
evolution. Because the rest wavelength of the \Lya line is in the far
ultraviolet, most DLA programs have focused on high$-$redshift systems
($z \gtrsim 2$) that can be detected from the ground, and at those
redshifts it is very challenging to study the DLA origin and
environment. At low redshifts,
studies have found a wide range of galaxies associated with DLAs both in terms of morphology and luminosity
\citep{ber_boi91,steidel95,lebrun97,bowen01,turn01,chen_lan03,rao03},
but the sample of low$-z$ galaxies studied so far remains small.
On the other hand, low$-$redshift 21 cm {\it emission} studies have shown
that gas with $N$(\ion{H}{1}) $> 2 \times 10^{20}$ cm$^{-2}$ can be
found in a variety of contexts \citep[e.g.,][]{hibbard01} ranging from
normal gas disks of high- and low-surface brightness galaxies to
dynamically disturbed structures such as tidally stripped gas spurs
and bridges or even detached intergalactic clouds. 
Galaxy interactions are expected to be more common at high redshifts,
so an even larger proportion of high$-z$ DLAs could originate in
disturbed structures. Interpretation of, e.g., the kinematics of DLAs,
could be confusing if the samples are composed of a mixture of these
different types of objects.

This raises a question: are there absorption signatures that can be
used to distinguish gas in different environments such as quiescent
galaxy disks vs. dynamically disturbed extraplanar structures such as
tidal tails or ram-pressure detritus?  All can have $N$(\ion{H}{1}) $>
2 \times 10^{20}$, but how do their physical characteristics compare?
Spectroscopic {\it absorption} studies of the \ion{H}{1} 21 cm line
provide unique information to address this question for several
reasons: (1) In the UV/optical, QSOs are effectively point sources,
but in the radio continuum, radio-loud QSOs are often significantly
extended and resolved \citep[e.g.,][]{lazio09}, and the absorption
structure can be mapped against the extended background QSO continuum
to obtain information on the spatial extent and characteristic sizes
of the clouds that comprise DLAs. In addition, radio interferometers
such as the Very Long Baseline Array (VLBA) can be employed to do this
mapping at much higher angular resolution than is possible in other frequency
 bands. Thus, radio observations provide unique constraints on very
small-scale structures in DLAs. Interestingly, the literature on 21cm
absorber sizes suggest vastly different sizes in different
environments.  For example, \citet{keeney05} argue that the 21 cm
absorption feature associated with tidal feature of NGC~3067 \citep{stocke91} shows similar profile and comparable strength over a scale of 20~milli-arcsecond. The authors suggest that the feature arises in an atomic gas structure of physical size $>$~2-20~$h_{70}^{-1}$~pc. However, based on their \HI\ emission map and photoionization model they concluded that the absorbing cloud is uniform on a scale of  5~$h_{70}^{-1}$~kpc.  It is possible that large coherent \HI\ clouds can contain clumpy internal structures on much smaller scales. For instance, 
21~cm absorbers in the Milky Way have been shown to contain
 much smaller structures with sizes on the order of tens of AU
\citep[e.g.,][]{lazio09}. Although, these scales have not yet been
probed in extragalactic gas clouds; the non detection of these small clouds can be attributed to the fact that they have a small volume filling factor as noted by \citet{lazio09}.
Very little information is available regarding the absorbing
cloud sizes in other galaxies,  which is one of the
motivations for the study presented in this paper (see below). (2) The
hyperfine 21 cm transition has a very low transition probability, so a
substantial \ion{H}{1} column is required to detect the line, 
and surveys designed to search for 21 cm absorbers in radio-loud QSOs 
 automatically select systems that are DLAs or at least sub-DLAs.
Thus, 21cm absorption reveals the kinematics and physical
characteristics of the neutral gas with minimal confusion from
substantially ionized gas (which can confuse analyses based on metal
ions such as \ion{Si}{2} or \ion{S}{2}). (3) Unlike 21 cm emission, 21
cm absorption depends on the spin temperature of the gas, and combined
with the high spectral resolution available with radio spectrometers,
this can provide valuable constraints on the physical conditions of
the absorber. (4)~\ion{H}{1}  21~cm absorption is not affected by dust and hence, such a study can address whether optical DLA studies are biased by dust extinction \citep[see,
e.g.,][]{ellison09}.

\begin{figure*}
\includegraphics[angle=-90,trim = 40mm 40mm 35mm 0mm, clip, width=7in]{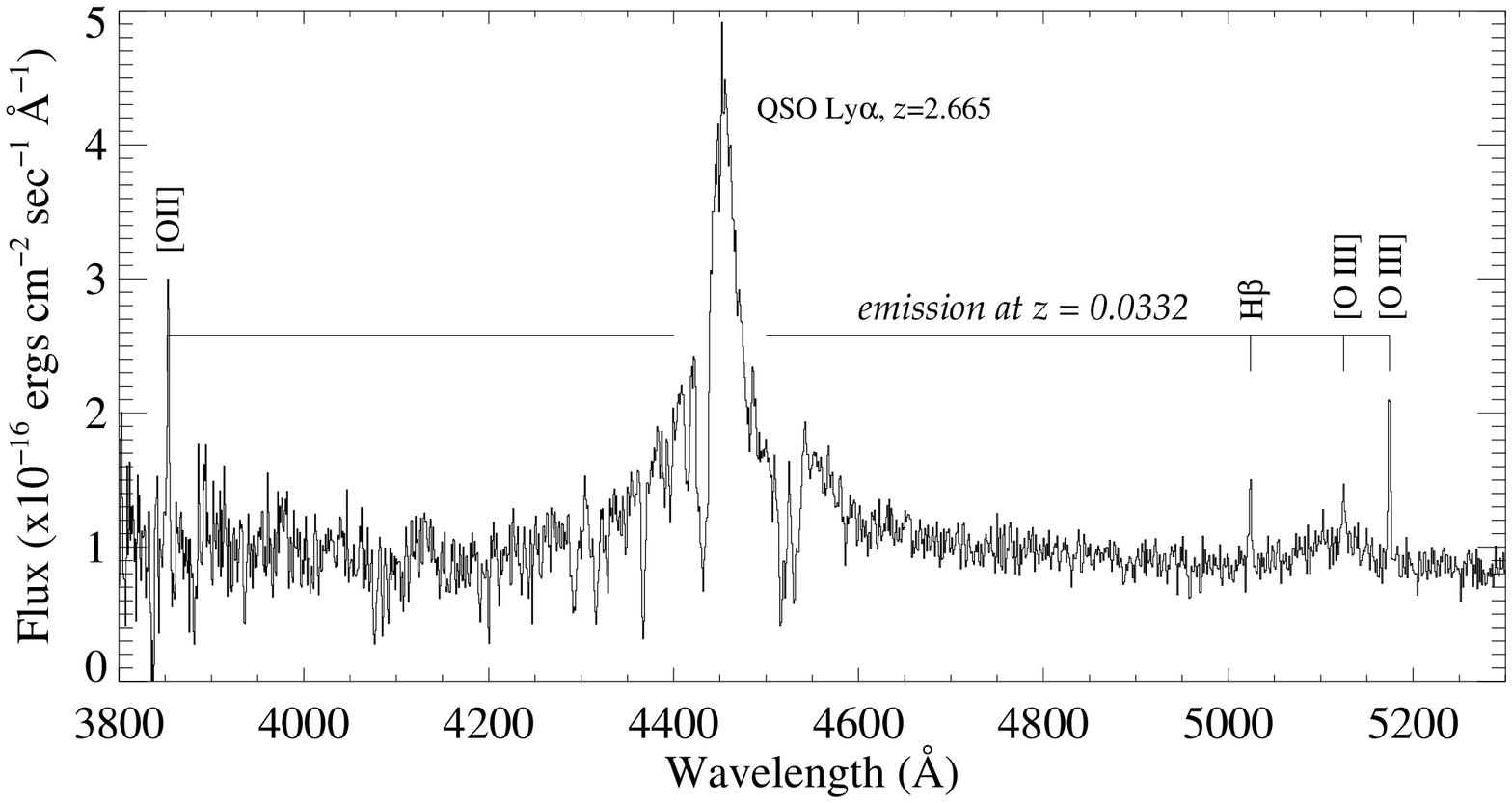}\\
\includegraphics[angle=-90,trim = 60mm 40mm 35mm 0mm, clip, width=7in]{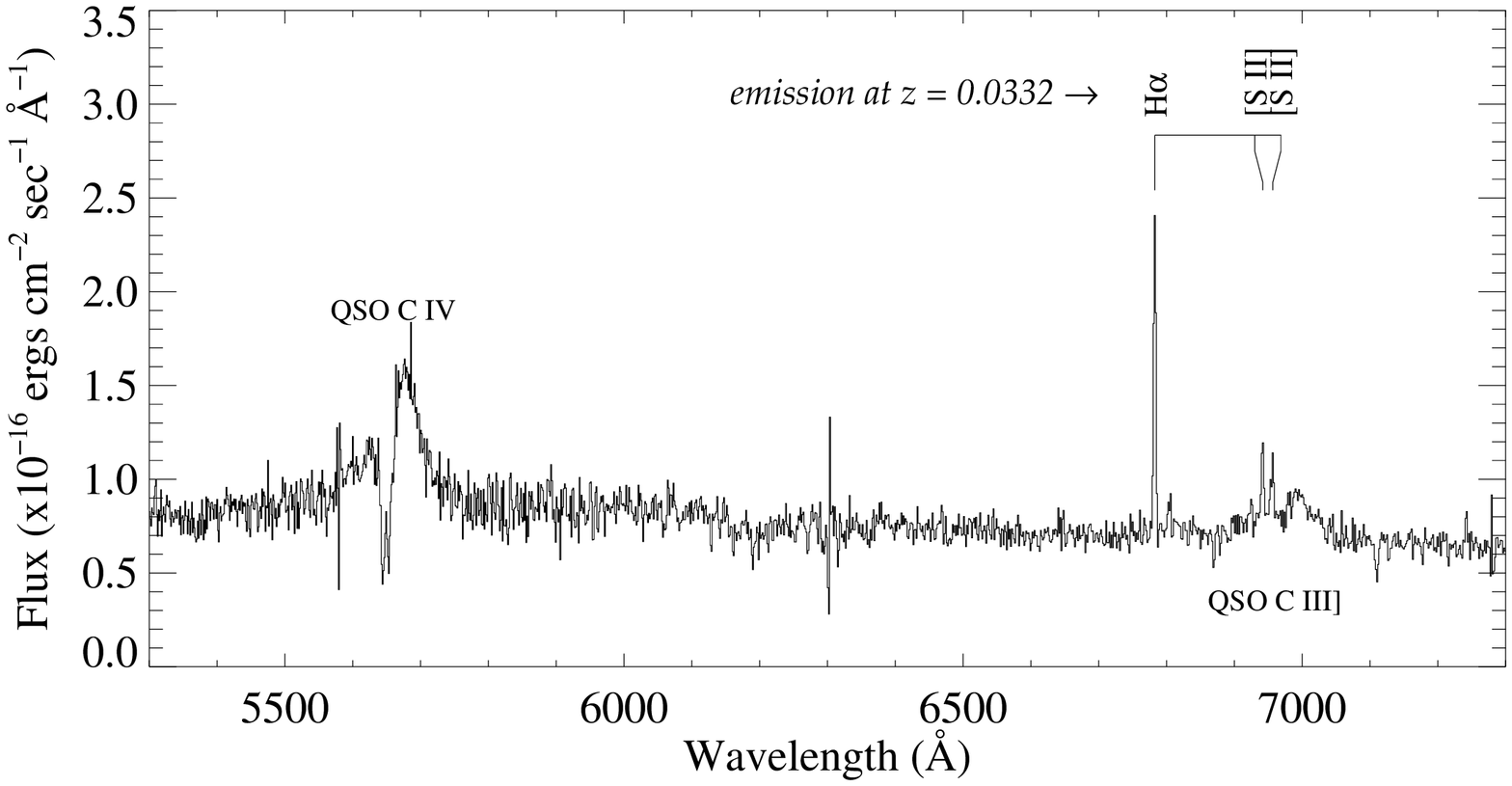}\\
\caption{\label{fig_spec} The SDSS spectrum of the $z=2.665$ QSO  \qsoname . 
While the broad \Lya, C~IV and C~III] emission  lines are seen at the redshift of the QSO, superimposed on the spectrum are narrow emission lines of [O~II], H$\beta$, [O~III], H$\alpha$ and [S~II] at a redshift of $z=0.0332$. The BAL nature of the QSO can be seen by the blue-shifted absorption near the QSO emission lines of \ion{C}{4}, \ion{N}{5} and possibly Lyman~$\alpha$. The complex features near 5585~\AA\ and 6300~\AA\ are artifacts of poor night sky subtraction.} 
\end{figure*}

For these reasons, various groups have applied 21 cm absorption
spectroscopy to the study of DLAs \citep[e.g.,][and references
therein]{kanekar04,kanekar05,curran07}.  Unfortunately, 21 cm
absorbers are rare, and blind surveys for 21 cm absorption are very
inefficient with current radio telescopes, so most previous studies
have searched for 21 cm absorption in known DLAs and thus are
dominated by high$-z$ systems.
 For example, \citet{macdonald09} searched 243 possible sources for 21~cm absorbers with $N$(\ion{H}{1}) $> 2 \times 10^{20}$ cm$^{-2}$ and 3282 possible sources with $N$(\ion{H}{1}) $> 2 \times 10^{21}$ cm$^{-2}$ in the Arecibo Legacy Fast Arecibo L-Band Feed Array (ALFALFA) Survey. They re-detected one previously known intrinsic \HI\ absorber, but no additional lines were identified.
   Likewise, searches for 21cm absorption
in \ion{Mg}{2} systems \citep{gupta09,kanekar09a} have mostly found
cases at relatively high redshifts where it is difficult to examine
the environment of the absorbers.  

To overcome this problem and assemble a sample of very low-redshift 21
cm absorbers suitable for environment studies, we have initiated a new
survey that uses a different strategy to find the 21 cm systems.
Our strategy to find low impact parameter sightlines through nearby galaxies is twofold. Firstly, we use the Sloan Digital Sky Survey \citep[SDSS,][]{york00} to directly search for background QSOs at small impact parameters to low$-$ redshift galaxies. Secondly, we search for QSO spectra that show emission lines from \emph{both} the QSO itself and a forground galaxy. From this sample, we have selected radio-loud quasars based on data from
the FIRST survey \citep{becker95} or the NRAO Very Large Array (VLA)
Sky Survey \citep{condon98}. The first part of our survey is to
observe the radio-loud QSOs with the Green Bank Telescope (GBT) to
determine if the foreground galaxy is a 21 cm absorber. Currently we are carrying out observations of 28 sight lines through 18 foreground galaxies from redshift 0.00154 to 0.2596 with projected distances varying from 1.7 to 109.7~kpc.
The second part is to follow up the GBT detections with higher angular resolution observations with either the VLA or the VLBA.

In this paper we present the results of the pilot study that we
conducted for this survey.  This initial program successfully detected
21 cm absorption from a foreground galaxy, and we have followed up
this detection with high-resolution VLBA observations and optical
imaging. The paper is organized as follows. We first present the
spectroscopic technique used for identifying foreground galaxies from
the spectra of background QSOs (\S~2). This is followed by a detailed
analysis of a QSO-galaxy pair identified using our technique and the
discovery of a 21 cm absorber in the foreground galaxy.  The optical
imaging and foreground galaxy and environmental properties are presented in \S~3, along with estimates of the foreground galaxy's star formation rate, metallicity and stellar mass. In \S~4 and \S~5 we describe our radio observations and analyze the properties of the 21 cm \ion{H}{1} absorber respectively. We discuss the physical nature and possible origin of the absorber in \S~6. Finally, in \S~7 we
summarize our findings. Throughout the paper we use $H_0 =
70~{\rm km~s}^{-1}~{\rm Mpc}^{-1}$, $\Omega_m = 0.3$, and $\Lambda_0 =
0.7$.

\section{Galaxy Discovery and Redshift \label{gal_discovery}}

The presence of a galaxy close to the line of sight of the quasar
\qsoname\ ($z_{\rm QSO} = 2.66521$) was discovered serendipitously by
one of us (DGY) during the compilation of QSO absorption-line catalogs
from the SDSS spectroscopic database \citep{york06b,lund09}.
 In addition to the usual broad
emission lines at $z=2.665$, the SDSS spectrum of the QSO show
narrow emission lines of [O~II], H$\beta$, [O~III], H$\alpha$, and
[S~II] at a much lower redshift. Figure~\ref{fig_spec} shows the SDSS
spectrum of \qsoname; the narrow emission lines of the foreground
galaxy are readily apparent. 
The RA and DEC of the galaxy, which is not cataloged in SDSS, is given in Table 1. The QSO and galaxy have very similar co-ordinates of course, and to distinguish between the two in this paper, we will use the SDSS designation for the QSO, but will refer to the galaxy as ``\galaxyname'', where GQ is an abbreviated version of ``galaxy on top of QSO''.
Such QSO-galaxy pairs at small impact 
parameters are generally not recognized as two distinct objects by the
SDSS target classification algorithms. Indeed, QSO-galaxy pairs at
small impact parameters are hard to find by any technique.  A recent
search for ``composite'' Sloan QSO spectra like the one shown in
Figure~\ref{fig_spec} (i.e, including emission features of lower$-z$
foreground galaxies) has unearthed 21 pairs with impact parameters
$\lesssim$ 10 kpc \citep{quash08,noter09b} out of $\approx 74,000$ QSO
candidates examined.  While such pairs are rare, they deserve special
attention for the purpose of studying the interstellar media of the
foreground galaxies.

Gaussian profile fits to the seven detected narrow (foreground)
emission lines in Figure~\ref{fig_spec} give an average redshift of
$z= 0.0332\pm0.0001$, with the error derived simply as the standard
deviation in the seven measurements. As SDSS spectra are always corrected
to heliocentric velocities, this redshift corresponds to a
heliocentric velocity of $9962 \pm 33$~\kms . After recognition of the
composite nature of the QSO spectrum, a simple visual inspection of
the SDSS images of the QSO immediately revealed a galaxy close to the
QSO line of sight, confirming the notion that a galaxy was
contributing light to the spectrograph fiber that had been centered on
the QSO.

As part of a follow-up investigation to search for interstellar Ca~II
and Na~I absorption lines arising from gas in the foreground galaxy,
we observed the QSO using the Dual Imaging Spectrograph (DIS) on the
Astrophysical Research Consortium (ARC) 3.5~m telescope at the Apache
Point Observatory (APO) on 06 January 2006. The DIS was configured with the
high-resolution gratings and a 0.9 arcsec slit, giving a spectral
resolution of $\simeq 80$~\kms\ (FWHM) at 6700~\AA . Although
conditions were too poor to obtain spectra with sufficient S/N to
detect absorption lines from the foreground galaxy, the H$\alpha$
emission line was detected, permitting an independent measurement of
the galaxy's redshift, at a resolving power of about twice that of the
SDSS data. We extracted spectra over a 5 arcsec aperture centered on
the QSO and tied the zero-point of the wavelength calibration to four
narrow skylines which were also covered by the observations. From the
H$\alpha$ line alone, we measured a redshift of $z=0.03321\pm
0.00003$, with the error derived from the standard deviation in the
set of differences between the rest wavelengths of the four sky lines
and the wavelengths we measured. After correcting the observed radial
velocity to a heliocentric one, we found that the H$\alpha$ emission
line gave a systemic velocity of $9932\pm 10$~\kms.

\begin{figure}
\centerline{\includegraphics[angle=0,scale =0.5]{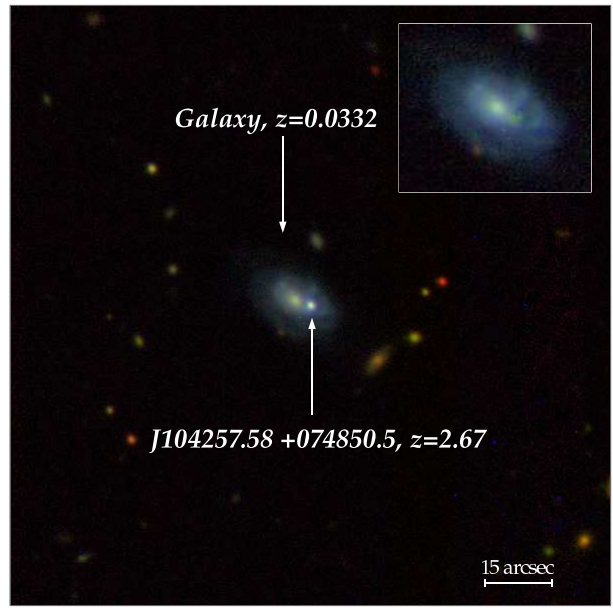}}
\caption{\label{fig_colorpic1} A color composite of $g$-, $r$- and
  $i$-band images of the field around \qsoname\ taken with the SOAR
  telescope. NE is to the top left of the image, and the scale is
  indicated bottom-right. The inset shows the galaxy after subtraction
  of the QSO profile.}
\end{figure}

\section{Optical Properties and Environment of the Foreground Galaxy \label{gal_optical}}

\subsection{SOAR Optical Imaging  \label{soar}}

Although the galaxy is clearly visible in SDSS images, little can
be inferred from the SDSS photometry due to the blending of the galaxy's image with that of the QSO.
 To learn more about the nature
of the foreground galaxy, we obtained deeper images with better
angular resolution using the SOAR Optical Imager
\citep[SOI,][]{schwarz04} on the 4.1~m Southern Astrophysical Research
(SOAR) telescope at Cerro Pach\'{o}n in Chile. Exposures were made
using $g$-, $r$-, and $i$-band filters for 10, 30, and 15 minutes, on
2 March, 28 February, and 1 March 2009, respectively.  The data were
processed in the conventional way, and calibrated astrometrically
using the {\tt SCAMP} software package (Bertin 2006). Individual
frames were co-added using the {\tt SWARP} software (Bertin et al
2002), which resamples and co-ads the individual images with the derived astrometric solution based on SDSS astrometry. The final coadded $g$-, $r$- and
$i$-band images had resolutions of 1.2, 0.7 and 0.8 arcsec FWHM,
respectively.  These were assembled into a color image following the
prescription given by \citet{lupton04}, using the data at the original
resolutions, and applying an inverse hyperbolic sine ($asinh$) scaling
algorithm. The resulting multicolor image of the QSO-galaxy pair is
shown in Figure~\ref{fig_colorpic1}.

\subsection{Photometry and Surface-Brightness Profile\label{sec:foregroundgal_phot}}

To derive accurate measurements for the physical parameters of
\galaxyname, we constructed a Point Spread Function (PSF) from a 2-D
Moffat profile fit to stellar images in each of the $g$-, $r$-, and
$i$-band images, and subtracted a suitably scaled PSF from the QSO
profile. The galaxy, with the QSO profile removed, and again formed
from a co-addition of all three colors, is shown inset in
Figure~\ref{fig_colorpic1}.

\begin{figure}
\centerline{\includegraphics[angle=90,scale=0.55]{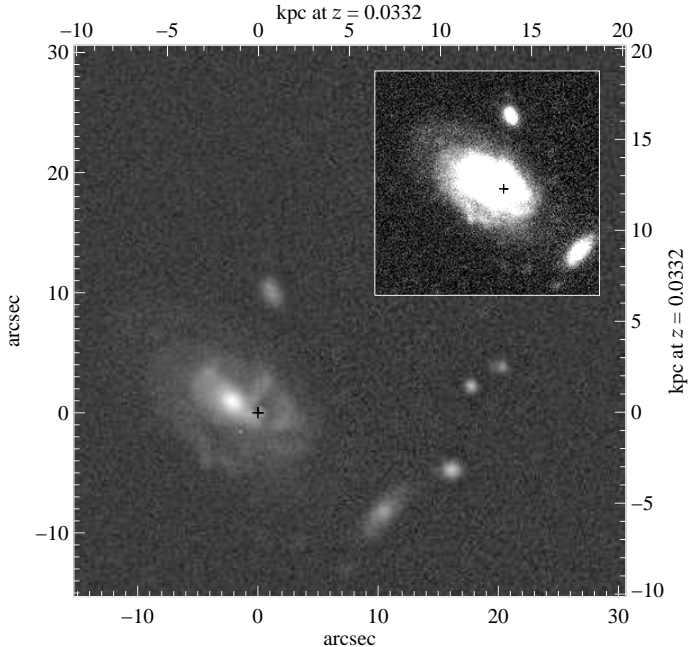}}
\caption{\label{fig_grscale} Grayscale representation of the r-band
 image of the galaxy GQ1042+0747 in front of
  \qsoname . The QSO profile has been removed, although its original
  position is labelled with a cross, and the image has been processed
  with a gentle unsharp mask. {\it Inset:} The
  same image of the galaxy reproduced using a much harder stretch
  in order to show the outer spiral arm to the NE of the galaxy. The
  position of the QSO prior to removal is again marked with a cross.}
\end{figure}

\begin{figure}
\centerline{\includegraphics[scale=0.55,angle=0]{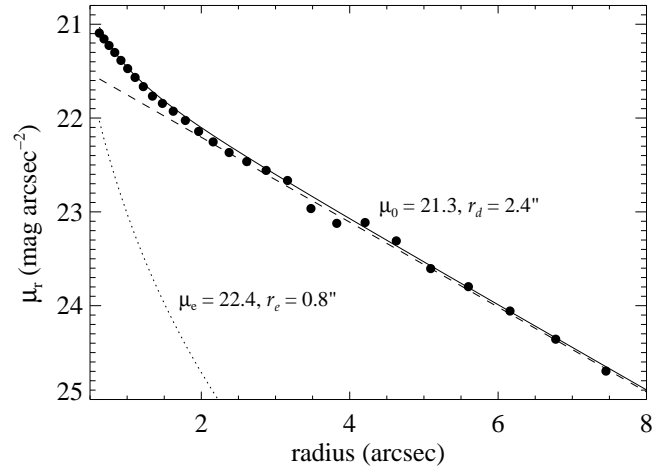}}
\caption{ The $r$-band surface brightness profile of
  the galaxy in front of \qsoname. The black dots are data points obtained using the IRAF isophote package. The final fit (solid line) is the result of co-adding an exponential disk profile (dashed line) and an
$r^{1/4}$ profile (dotted line). The relevant parameters for each
model are given alongside the profiles. }
\label{fig_sb}
\end{figure}

\begin{table*}
\caption{Properties of \galaxyname, the Foreground Galaxy in the QSO-Galaxy Pair  \label{tbl-gal_param}}
\centerline {
\begin{tabular}{ll }
\hline\hline
Center RA and DEC       \dotfill  &  10:42:57.74 +07:47:51.3\\
Redshift $z_{\rm{gal}}$ \dotfill  &  $0.03321\pm 0.00003^a$,  $cz = 9932\pm 10$~\kms\\
Impact parameter        \dotfill  &  2.5$''$ $\equiv$ 1.7 kpc at $z_{\rm{gal}}$\\
Magnitudes$^b$          \dotfill  &  $g=18.59$, $r=18.05$, $i=18.08$\\
MW extinction           \dotfill  &  $g=0.12$, $r=0.08$, $i=0.06$\\
Absolute magnitudes$^c$ \dotfill  &  $M_g = -17.35$, $M_r = -17.85$, $M_i = -17.80$\\
Major/Minor axis ratio  \dotfill  &  1.55 \\
Inclination$^d$         \dotfill  &  50 degrees\\
Surface Brightness ($r-$band)      \dotfill  &  $\mu_o = 21.3$, $r_d = 2.4''$; $\mu_e = 22.4$, $r_e = 0.8''$  \\
Log Metallicity         \dotfill  & $-0.27\pm 0.05$ (N2 index); $-0.32\pm 0.03$ (O3N2 index) \\
\hline
\multicolumn{2}{l}{\footnotesize $^a$  Redshift is from APO spectrum
  of H$\alpha$ line.}\\
\multicolumn{2}{l}{\footnotesize $^b$  Isophotal magnitudes from {\tt
    sextractor}. Formal errors on these values are $=\pm0.02$ mags.}\\
\multicolumn{2}{l}{\footnotesize $^c$ Corrected for Milky Way
  extinction, but no k-correction applied.} \\
\multicolumn{2}{l}{\footnotesize $^d$ cos$^{-1}$($b/a$), where $a$ and
  $b$ are the major and minor isophotal axes.}\\
\end{tabular}}
\end{table*}

A more detailed representation of the galaxy in the $r$-band image is
shown in Figure~\ref{fig_grscale}. To better highlight the small-scale
structure in the galaxy, we used the technique of unsharp masking,
using the procedure outlined in \citet{jenkins05}. We found that
``gentle'' unsharp masking using a Gaussian with width $\sigma = 5$
pixels most effectively enhanced the low-level morphological features
of the galaxy.  In Figure~\ref{fig_grscale}, this unsharp mask image
is shown, again scaled using the $asinh$ scaling algorithm. The image
shows that the galaxy is an inclined spiral, perhaps with a bar at its
center. Some evidence of spiral arms can be seen to the SW of the
center (below and to the right of the QSO's position). A much harder
stretch of the original image (the non unsharp mask data), shown inset
in Figure~\ref{fig_grscale}, clearly reveals an outer spiral arm to
the NE of the galaxy (top left of the inset). The main image in
Figure~\ref{fig_grscale} also shows a `plume', some $2-3$ arcsec
directly above the position of the QSO, which is not an artifact of
the QSO profile sutraction. Determining whether this is simply part of
the normal spiral pattern of a galaxy, or, perhaps related to what
might be a companion dwarf galaxy, SDSS J104257.52+074900.5, $\sim 10$ arcsec due north of the
QSO's position, will require higher resolution data. The companion dwarf galaxy has magnitudes measured from the SOAR data of
$g=22.1$, $r=21.5$ and $i=21.2$, with errors of 0.1 mags.

With these images, we were able to measure the galaxy properties
listed in Table 1. The magnitudes of the galaxy and its major-to-minor
axis ratio were derived using the {\tt sextractor} software package
\citep{sextractor}; the extinction from dust in the Milky Way was
taken from the maps of \citet{schl98}, and these values were used to
correct the absolute magnitudes (although no $k$-correction was
applied).

To derive a surface brightness profile, we used the {\tt ellipse}
routines in the STSDAS package {\tt isophote} \citep{Jedrz87} to fit
the $r$-band data. The results are shown in Figure~\ref{fig_sb}. 
The profile is that observed along the semi-major axis --- no other
corrections have been applied, and no attempt was made to first
deconvolve the data using the PSF. As Figure~\ref{fig_sb} shows, more
than a simple exponential disk model is required to explain the data,
so we fit a combination of a disk model with a classical $r^{1/4}$
profile. When the intensity profile of a galaxy is converted to
surface brightness, $\mu$, the profiles are given by:

\begin{eqnarray}
\mu_1 & = & \mu_e + 8.325 [ (\frac{r}{r_e})^{1/4} -1 ]\\
\mu_2 & = & \mu_0 + 1.086\:\frac{r}{r_d}
\end{eqnarray}

\noindent
where $r_e$ and $r_d$ are the scale lengths of the bulge and disk
components respectively, and $\mu_e$ and $\mu_0$ are the surface
brightness values at those radii. We fitted a combination of these
theoretical profiles to the observed values of $\mu$ by minimizing the
value of $\chi^2$ between the fit and the data, and derived the values
listed in Table~\ref{tbl-gal_param}. The resulting fit is shown in
Figure~\ref{fig_sb}.

\subsection{Star-Formation Rate, Metallicity, and Stellar Mass \label{sec:foregroundgal}}

\galaxyname\ is a low luminosity spiral with $L = 0.048~L_{\star}$ in
the r-band  \citep[using M$_r^*$=$-$21.2 from Table~2 of][]{blanton03}. Emission-line measurements from the SDSS spectrum of the
foreground galaxy are presented in Table~\ref{tbl-emlines}. The 3 arcsec-diameter fiber used to make these  measurements was centered at the QSO's position. Most of the observed region covered the disk; some
parts of the central region of the foreground galaxy were also
captured. We estimated the star formation rate (SFR) from the H$\alpha$
\citep{kenn98} and O[II] transitions \citep{kewley04} using the
following relationships,
\begin{equation} \label{eq-sfr_h}
{\rm SFR(H\alpha)}=7.9 \times 10^{-42} \times {\rm L(H\alpha, ergs~s^{-1})~}M_{\odot}~{\rm yr^{-1}} 
\end{equation}
\begin{equation}\label{eq-sfr_o}
{\rm SFR(O[II])}=6.58 \times 10^{-42} \times {\rm  L(O[II], ergs~s^{-1})~}M_{\odot}~{\rm yr^{-1}}
\end{equation}
The SFR  was found to be 1.26$\times$10$^{-2}$ and
8.6$\times$10$^{-3}~M_{\odot}$~yr$^{-1}$ using equation~\ref{eq-sfr_h} and \ref{eq-sfr_o} respectively, over an area of 3.1~kpc$^2$ in the rest frame of the \galaxyname. This
corresponds to SFR surface density of 4.1$\times$10$^{-3}$ and
2.8$\times$10$^{-3}$~M$_{\odot}$~yr$^{-1}$~kpc$^{-2}$, respectively. The
Balmer decrement, I (H$\alpha$/H$\beta$)= 3.0$\pm$0.4 suggests a low
dust content in this galaxy. The metallicity of the galaxy estimated
using the N2 index \citep{pettini04} is log (O/H) + 12 =
8.42$\pm$0.05. Assuming a solar abundance of log(O/H$_{\odot}$)=8.69
\citep{asplund09}, the logarithmic measured metallicity is then
$-0.27\pm0.05$ dex relative to solar. The O3N2 index yielded a similar
metallicity of $-0.32\pm0.03$ dex relative to solar.

\begin{table*}
\caption{Foreground-Galaxy Emission-Line Measurements and Implied Star-Formation Rates \label{tbl-emlines}}
\begin{tabular}{lcccll}
\hline\hline
Line\tablenotemark{a}  & EW$_r$ & I$_r$   & L$_r$   &  SFR surface density  \\
 & (\AA) & ($\times$10$^{-16}$~ergs~cm$^{-2}$~s$^{-1}$)  &  ($\times$10$^{39}$~ergs~s$^{-1}$)  & ($M_{\odot}$~yr$^{-1}$~kpc$^{-2}$) \\
\hline
H$\alpha$\dotfill       			       &8.8$\pm$0.3 & 6.7$\pm$0.2            & 1.6$\pm$0.06     & 0.0041 \\
$[$\ion{O}{2}$]$\dotfill 		       &5.1$\pm$0.6 & 5.3$\pm$0.6            & 1.3$\pm$0.2 	   & 0.0028\\
H$\beta$\dotfill         			       &2.5$\pm$0.4 & 2.2$\pm$0.3            & 0.56$\pm$0.08   & n/a\\
$[$\ion{N}{2}$]\lambda$6585\dotfill &1.3$\pm$0.3	 & 0.96$\pm$0.18  &0.24$\pm$0.06	   & n/a\\
$[$\ion{O}{3}$]\lambda$5008\dotfill &4.7$\pm$0.3 & 4.1$\pm$0.3	      &1.0$\pm$0.07       & n/a\\
$[$\ion{O}{3}$]\lambda$4960\dotfill &1.5$\pm$0.3 & 1.6$\pm$0.3	      &0.4$\pm$0.07       & n/a	\\
\hline
\end{tabular}
\tablenotetext{a}{Vacuum wavelength}
\end{table*}

\begin{figure*}
\includegraphics[scale=0.85, angle=90]{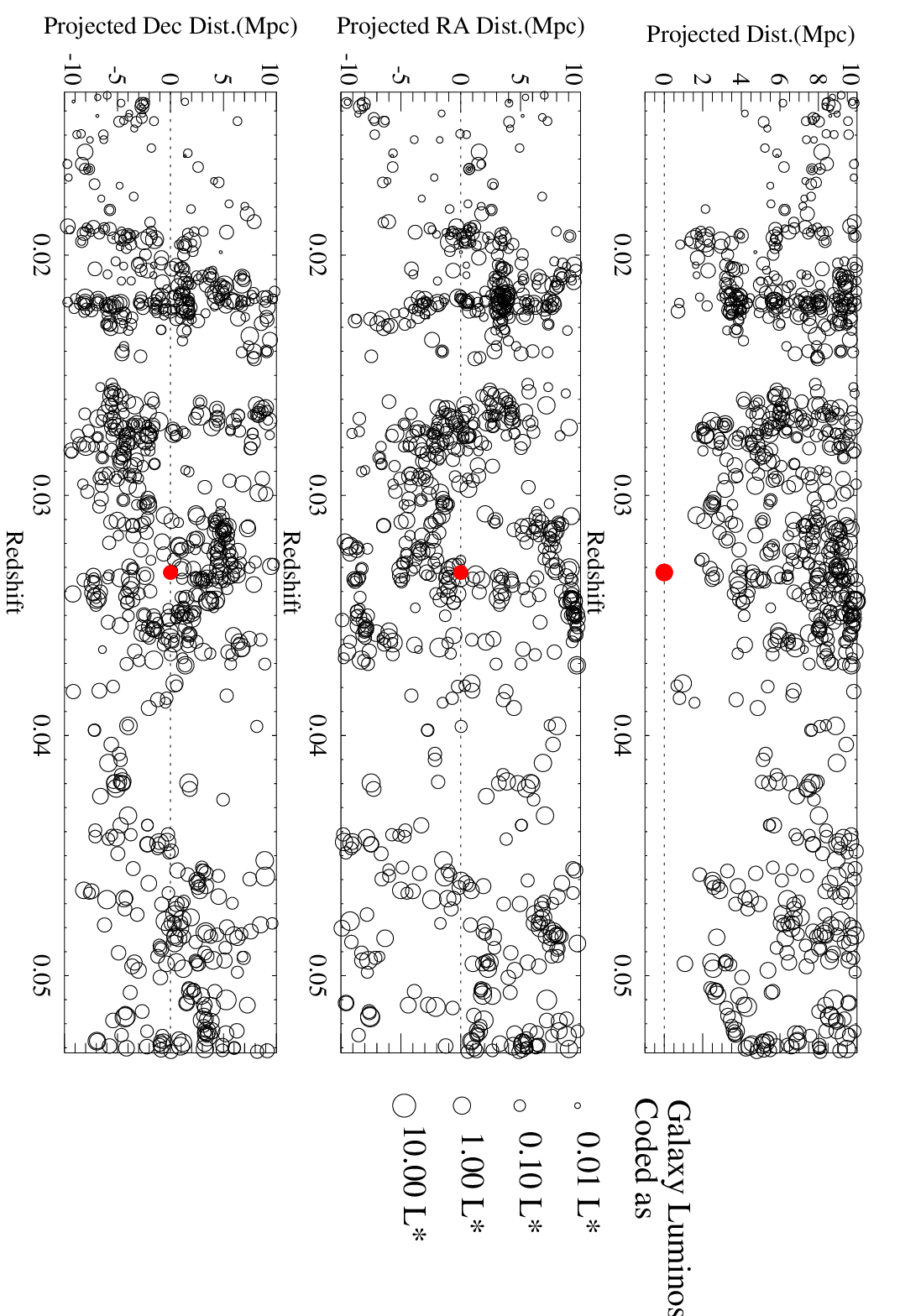}
\caption{ Projected distance of SDSS spectroscopic galaxies in
the field of \qsoname.  The sight line to the QSO is indicated with a
dashed line, and galaxies are plotted with open circles with size 
indicating the galaxy luminosity according to the legend at the
right. \galaxyname\  is plotted with a large red circle. {\it Top:}
Absolute projected distance from the sight line. {\it Middle:}
Projected distance in the RA direction. {\it Bottom:} Projected
distance in the declination direction. It is evident that the galaxy is relatively isolated and the nearest galaxy is $\sim$~2~Mpc away. The galaxy can be considered as a field galaxy although it lies at the edge of a large-scale structure filament.} 
\label{sdsscylinder}
\end{figure*}

\begin{figure}
\includegraphics[scale=0.52, angle=0]{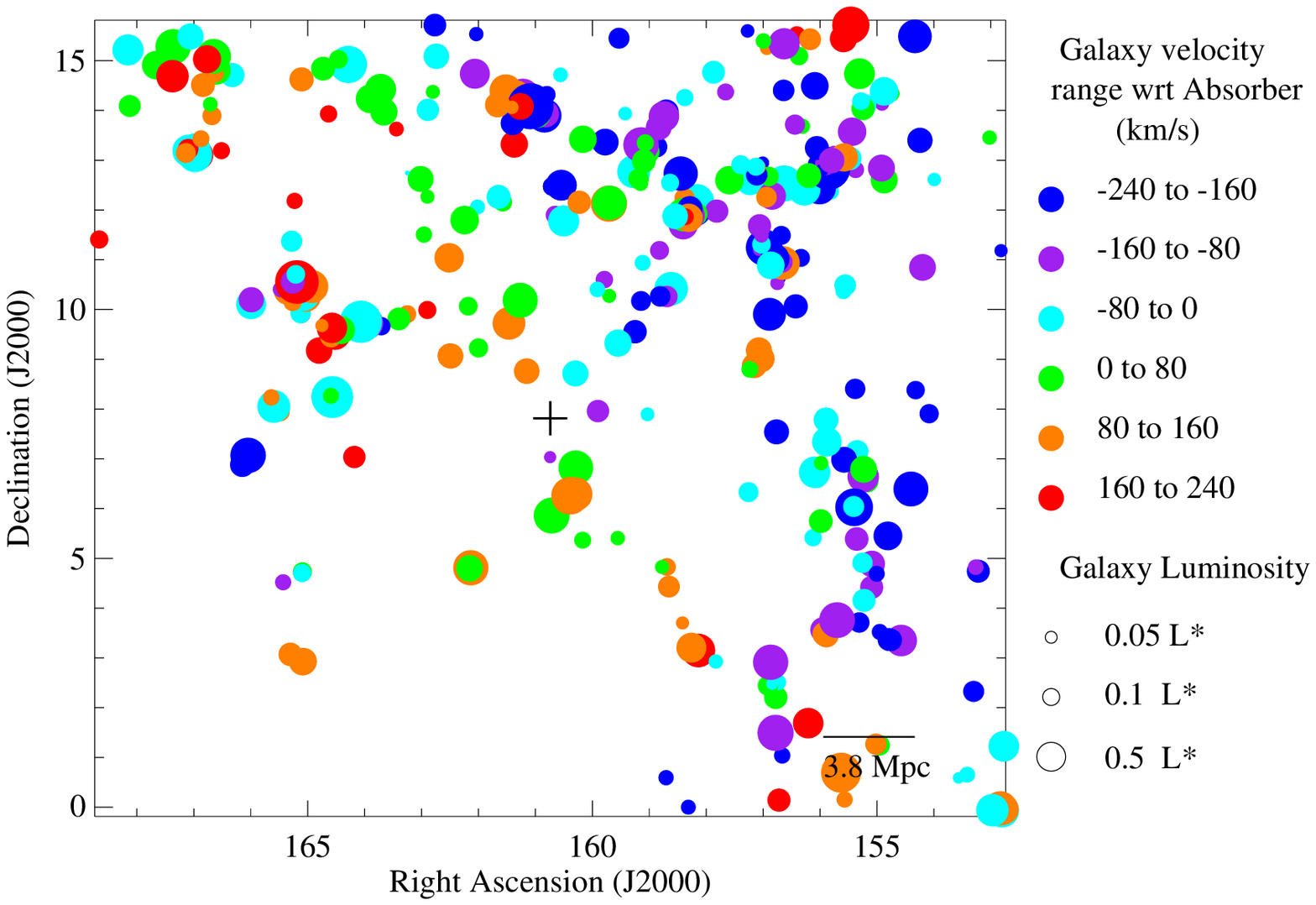}
\caption{\label{sdssslice} Distribution of SDSS galaxies (filled
circles) near the \qsoname\ sight line (plus symbol) in Right
Ascension and Declination.  Following the legend at upper right,
the colors of the circles indicate the velocity difference between the
galaxy and the centroid of the 21cm absorption detected in \galaxyname
, ranging from $v_{\rm galaxy} - v_{\rm 21 cm} = -240$ to +240 km
s$^{-1}$. The circle size represents the galaxy luminosity as shown in
the lower right key. The plot clearly shows a density gradient  between the northwestern and the southeastern side of \galaxyname\ confirming that it relatively isolated galaxy at the edge of a large scale structure filament. }
\end{figure}

It is now well-known that galaxies in the nearby universe follow
mass-metallicity and luminosity-metallicity relationships
\citep[e.g.,][]{tremonti04}.  It is often argued that the
mass-metallicity trend has important implications regarding galactic
outflows, so it is of interest to consider where \galaxyname\ is
located in the mass-metallicity trend. Referring to Figure 5 in
\citet{tremonti04}, we see that \galaxyname\ is a relatively
low-luminosity galaxy and has a somewhat low metallicity compared to
other SDSS galaxies of comparable absolute magnitude, but given the
scatter in the relationship, the observed luminosity and metallicity
of \galaxyname\ are in good agreement with the usual
luminosity-metallicity trend.  To estimate the stellar mass of the
galaxy, we use equation 1 of \citet{mcintosh08}, which is based on the
stellar {\it M/L} ratios from \citet{bell03}, and we obtain
\begin{equation}
{\rm log} \ \left( \frac{{\it M}_{stars}}{{\it M}_{\odot}}\right) \approx 9.1.
\end{equation}
Thus, the metallicity is low for its stellar mass [see
Figure 6 in \citet{tremonti04}], but again \galaxyname\ follows the
general trend within the observed scatter.  We will discuss the
implications of the mass and metallicity of the galaxy compared to its
absorption properties in \S~6.

\subsection{Large-Scale Environment of \galaxyname \label{sec:galenv}}
A benefit of using SDSS to select low$-z$ galaxy-QSO pairs is that
SDSS provides detailed information about the global context of the
galaxy and its affiliated absorption. We will show below that despite
the fact that \galaxyname\ is clearly a star-forming spiral galaxy, it
has a surprisingly low \ion{H}{1} column density in its inner disk,
and overall it has a low \ion{H}{1} mass. The cause for its \ion{H}{1}
deficiency is an interesting question, and we will hypothesize that
this is related to the galaxy's environment.  To set the stage for
this discussion, we show the large-scale distribution of SDSS galaxies
near \galaxyname\ in Figures \ref{sdsscylinder} and
\ref{sdssslice}. Figure~\ref{sdsscylinder} shows the distribution of
SDSS galaxies versus redshift along the line of sight to \qsoname\ out
to $z = 0.053$; the panels show the overall (absolute) projected
distance of the galaxies from the sight line (upper panel) and the
projected distance in the right ascension and declination directions
only (middle and lower panels, respectively). This is a cylinder cut
from SDSS spectroscopic data (up to DR7) centered on the QSO. SDSS is a
magnitude-limited survey, so the galaxy points are plotted with symbol
sizes that indicate the object's luminosity based on values from NYU Value Added Catalog \citep{nyuvagc}, as shown in the legend,
to give the reader a sense of the survey completeness for various
galaxy luminosities. \galaxyname\ is shown with a large red dot.  We
show a substantial range in redshift so that voids and large-scale
structures in the galaxy distribution can be visually recognized.  In
order to zoom in on the more immediate vicinity of \galaxyname,
Figure~\ref{sdssslice} shows the RA and Dec of only galaxies with
redshifts within $\pm$240 km s$^{-1}$ of \galaxyname . In this figure,
the symbol color indicates the velocity difference between the galaxy
redshift and the redshift of the 21cm absorption detected in
\galaxyname\ (\S~4), and the symbol size indicates
the galaxy luminosity, as reflected in the figure legend.

From Figures~\ref{sdsscylinder} and \ref{sdssslice}, we notice the
following aspects of the \galaxyname\ environment: (1) The galaxy is
located near the boundary of a large-scale structure and most likely  in a low density region.
 This is readily apparent in Figure~\ref{sdssslice} --- there is
a paucity of galaxies southeast of \galaxyname, but many galaxies are
found to the northwest with a prominent galaxy group apparent at RA
$\approx 157^{\circ}$ and Dec $\approx 13^{\circ}$. (2) Qualitatively,
\galaxyname\ appears to be located within the boundaries of the
large-scale structure, but nevertheless it appears to be relatively isolated --- the nearest-neighbor galaxy is  $>$~1.8 Mpc away in projection and is
offset in velocity, so the three-dimensional nearest-neighbor distance
could be significantly larger.  Traditionally, \galaxyname\ would be
considered a ``field'' galaxy; in more recent nomenclature, the galaxy
might be referred to as a ``wall'' galaxy.  A detailed spectroscopic study of the immediate environments of this galaxy would help to clarify how isolated it is. We believe that the
relative isolation may have been an important factor that has affected this
galaxy's evolution, as we will discuss in \S~6.

\section{21 cm Absorption Spectroscopy \label{sec:radio_21cm}}

To search for 21 cm absorption from the ISM/halo of the foreground
galaxy, SDSS J104257.58 + 074850.5 was observed with three
complementary National Radio Astronomy Observatory (NRAO) telescopes:
the 100~meter GBT, the VLA in the B configuration, and the VLBA.  The optical
redshift of \galaxyname\ was used to choose the bandpasses for the
21~cm observations.  To maximize the effectiveness of our search for
21~cm absorption in the general vicinity of the foreground galaxy, the
 VLA was setup to provide the broadest spectral coverage
that could be afforded at a reasonable spectral resolution.  This
resulted in sub-optimal spectral resolution for the VLA (10.6~\kms ),
but was overcome by the excellent resolution of the GBT
(0.33~\kms ). Following the detection of the absorber, VLBA \HI
observations were then carried out centered at the redshift of the
absorber. Spectral resolution of 0.9~\kms was achieved in our VLBA
observations along with a superb milli-arcsecond (mas) spatial resolution of
22~mas $\times$~9~mas.

\begin{figure}
\includegraphics[angle=0,scale=0.65]{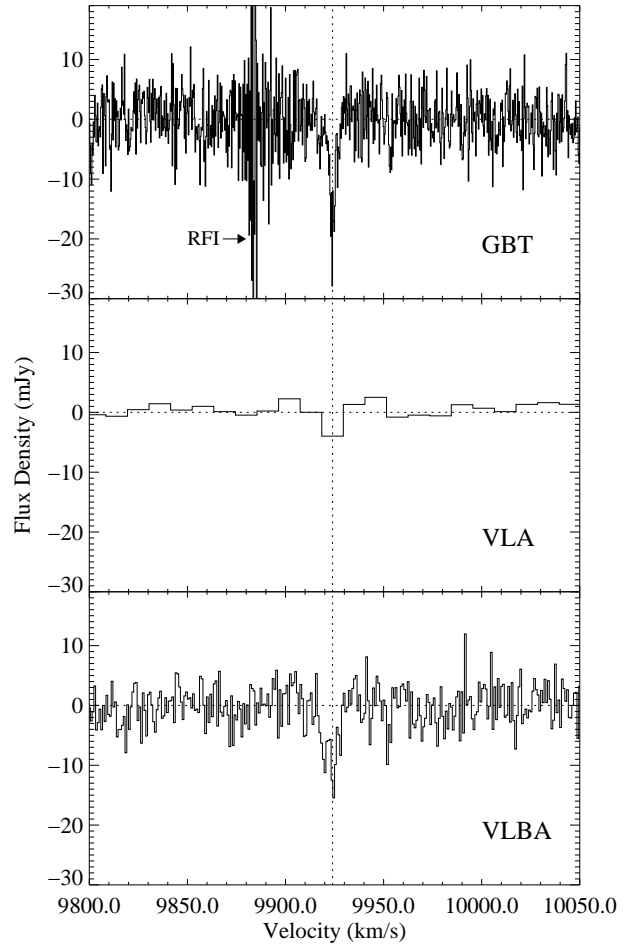}
\caption{Continuum-subtracted GBT, VLA, and VLBA spectra of \qsoname\ covering the velocity range of redshifted 21~cm absorption associated with the foreground galaxy. The VLBA spectrum is extracted from pixels with more than 400~$\mu$Jy (i.e 5~$\sigma$) continuum flux. The vertical dotted line marks a significant absorption feature that is evident in all the three spectra at 9924~\kms .  An expanded plot of the 21 cm absorption profile in the GBT data and its comparison with VLBA data is shown in Figure~\ref{j1042_gbt_zoom}.}
 \label{j1042_gbt_vla_vlba}
\end{figure}

\begin{figure}
\includegraphics[angle=0,scale=0.6]{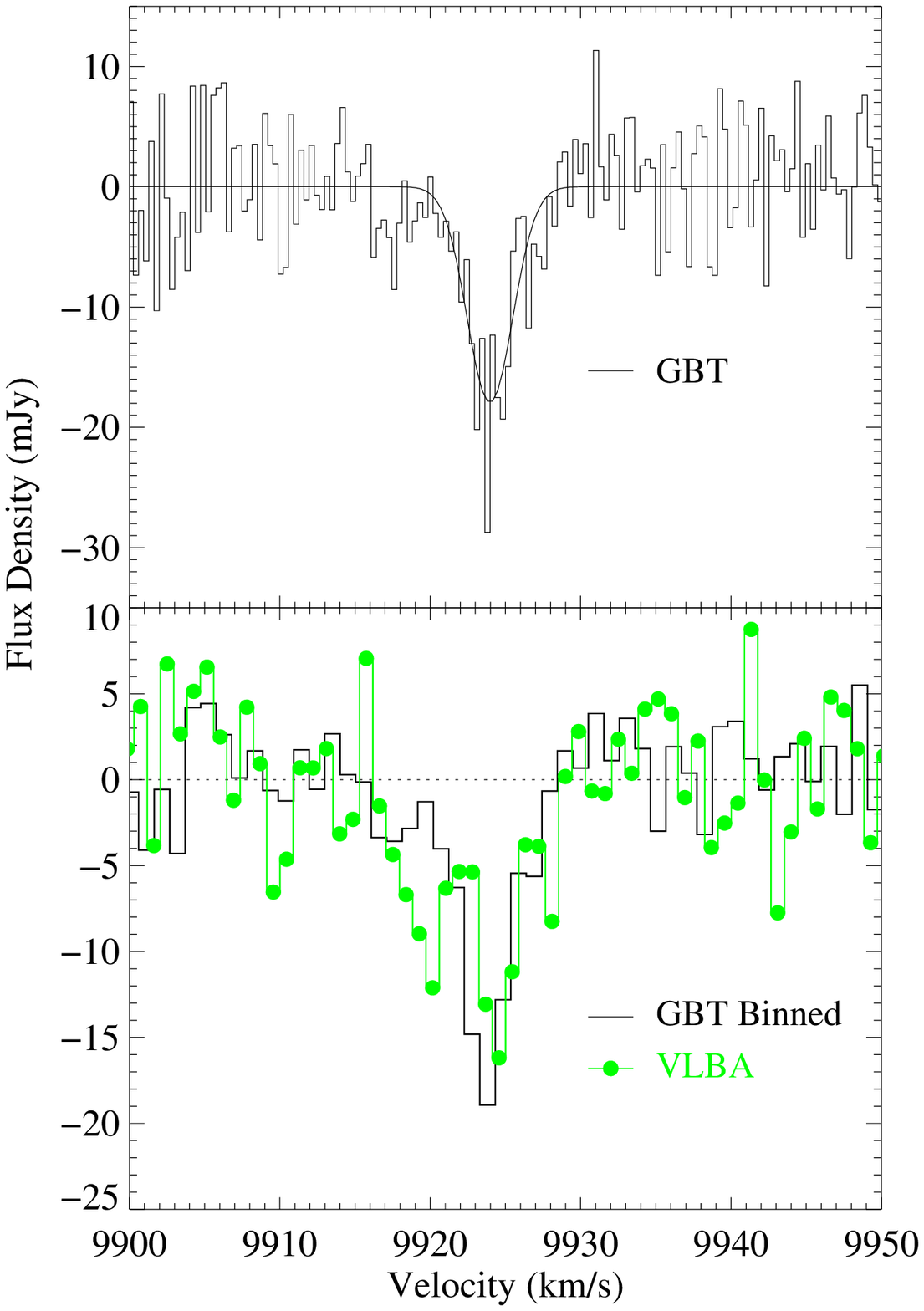}
\caption{{\it Top:} The continuum-normalized GBT spectrum of the 21 cm absorption detected at 9923.97~$\pm$~0.17~\kms in the spectrum of the background QSO, \qsoname. 
The absorption profile can be represented by a single Gaussian component with a FWHM of 3.6~$\pm$~0.4~\kms shown on the spectrum.  A second component (2.3~$\sigma$) can be seen centered at 9917.6~\kms. The properties of the absorber are consistent with that of a quiescent cloud that has a kinematic temperature of $ < 283$ K.  {\it Bottom:} A comparison between the GBT spectrum and the VLBA spectrum of the 21 cm absorption feature. The GBT spectrum is binned to match the resolution of the VLBA spectrum. There is a good match between the two spectra for the primary component, i.e towards the higher velocity-end of the feature. However, the VLBA spectrum shows a much stronger second component (4.6~$\sigma$) as compared to the GBT spectrum.
\label{j1042_gbt_zoom}}
\end{figure}

\subsection{Green Bank Telescope Observations \label{sec:obs_gbt}}

We observed \qsoname\ with the GBT for a total on-source integration
time of nearly 85 minutes over two sessions on 5-6 August 2006 as part of program GBT06B-052.  
The source was observed in two frequencies corresponding to the \ion{H}{1}
21~cm and OH 18~cm transitions at the redshift of the foreground
galaxy. Unfortunately, the OH data were corrupted by severe
interference and were rendered unusable, so hereafter we only discuss
the 21 cm data.  We used the dual polarization L-band system with a
bandwidth of 12.5~MHz. Nine-level sampling and two IF settings were
employed to provide 8196 channels with 1.56~kHz (0.33~\kms ) per
channel, covering a total velocity range of 2800~\kms . The source 3C
286 was used as the flux density calibrator, and a calibration of
1.65$\pm$0.05 K/Jy was determined from the mean of the four
independent recordings.  The observation was carried out in standard
position switching scheme by cycling through the ON-OFF
sequence, dwelling for 300 seconds at each position. Data was recorded every 30 seconds to minimize the affect of radio frequency interference (RFI). A position offset
of +20$^{\prime}$ in Right Ascension was adopted for the OFF position
so that presence of any confusing sources in the OFF position could be
tracked. The \ion{H}{1} data showed slight sinusoidal baseline
modulation on scales of $\approx$1000~\kms. However, most of the data
were uncorrupted on smaller velocity scales except for a few scans
which showed strong interference features. The final spectrum, shown
in Figure~\ref{j1042_gbt_vla_vlba} and \ref{j1042_gbt_zoom}, was
obtained by adding the uncorrupted data from our two observing
sessions using the NRAO software GBTIDL. The rms noise in the final
spectrum is 4.2~mJy.  We clearly detect 21cm absorption at
9924 km s$^{-1}$, i.e., very close to the optical redshift  (H$\alpha$) of the foreground galaxy, as
we discuss in \S~\ref{sec:components}. We note that the GBT data are corrupted by an RFI
feature at $\approx 9880$ km s$^{-1}$, so any absorption/emission near
that velocity could be difficult to detect, and in some regards it is
more effective to use the VLA or VLBA data to search for features
there.

\subsection{Very Large Array Observations  \label{sec:obs_vla}}
\begin{figure}
\includegraphics[ angle=0,scale=0.32]{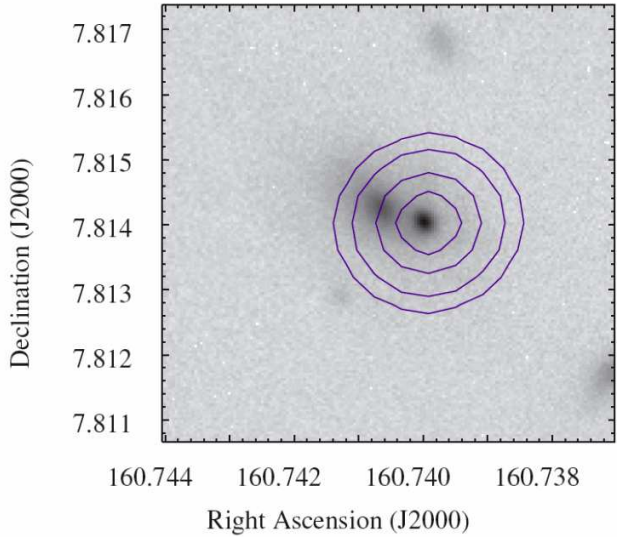}
\caption{The VLA B-array 1.4~GHz continuum image of \qsoname (contours) overplotted on the SOAR $i-$band image (grayscale). The resolution achieved in this image is 6$^{\prime\prime}$. The contours are plotted at 50, 100, 200, 300, and 380 mJy/beam. The quasar is unresolved, and the extent of the contours reflects the VLA beam size in this configuration.
\label{vlacontin}}
\end{figure}

\qsoname\ was observed for two hours with the VLA in
B-configuration on June 30, 2006 (project ID AT330). Only 22 antennas were available
because of the Expanded VLA (EVLA) conversion that was occurring at the
time.  The correlator was configured in the single polarization 2IF
mode with 64 spectral channels at a frequency resolution of 48.8 kHz
($\sim$ 10.6 kms$ ^{-1}$) per channel to cover a total bandwidth of
3.125 MHz ($\sim 650$ kms$^{-1}$).  The data were calibrated following
the standard VLA calibration and imaging procedure in the Astronomical
Image Processing Software (AIPS).  Absolute uncertainty in the
resulting flux density scaling is about 15\%, and this is the formal
uncertainty we quote for all physical parameters derived from the flux
density.

We show in Figure~\ref{vlacontin} the 1.4 GHz continuum image from the
VLA (contours) overplotted on the SOAR $i-$band image (grayscale).  In
this figure, the synthesized beam produced using natural weighting is
5.8$^{\prime\prime}$ $\times$ 5.5$^{\prime\prime}$ (PA=$-73^\circ$).
The 1.4 GHz continuum image shows a compact source, which is
unresolved by the VLA, centered at $\alpha(J2000)=10^h 42^m 41.58^s$
and $\delta(J2000)=+07^d 48^m 58.50^s$ with a peak flux density of
$395\pm59$ mJy.\footnote{The $1\sigma$ noise in the continuum image is
  about 0.2 mJy beam$^{-1}$, which reflects the dynamic range of the
  data rather than thermal noise.}  This is about 3\% larger than the
1.4 GHz flux density of the same source found in the archival VLA
FIRST Survey \citep{becker95}.  However, this difference is well
within the absolute calibration uncertainty of the VLA.  The
unresolved radio source is centered on the QSO and is clearly offset
from the center of the foreground galaxy by 2.5$^{\prime\prime}$ to
the southwest.

The continuum-subtracted VLA \ion{H}{1} spectrum of \qsoname\ covering
the velocity range between 9820 \kms and 10150 \kms is shown in
Figure~\ref{j1042_gbt_vla_vlba}.  The rms noise in each 10.6 \kms
channel maps is $\sim$0.7 mJy beam$^{-1}$ (21 K), and no \ion{H}{1} is
significantly detected in emission.  The narrow absorption feature
seen at $V=9924$ \kms is spectrally unresolved with an average optical
depth of $\tau_{\rm H~I}=0.0101\pm 0.0018$ ($5.6\sigma$) over
the 10.6 \kms channel width.

\begin{figure*}
\includegraphics[angle=0,scale=.8]{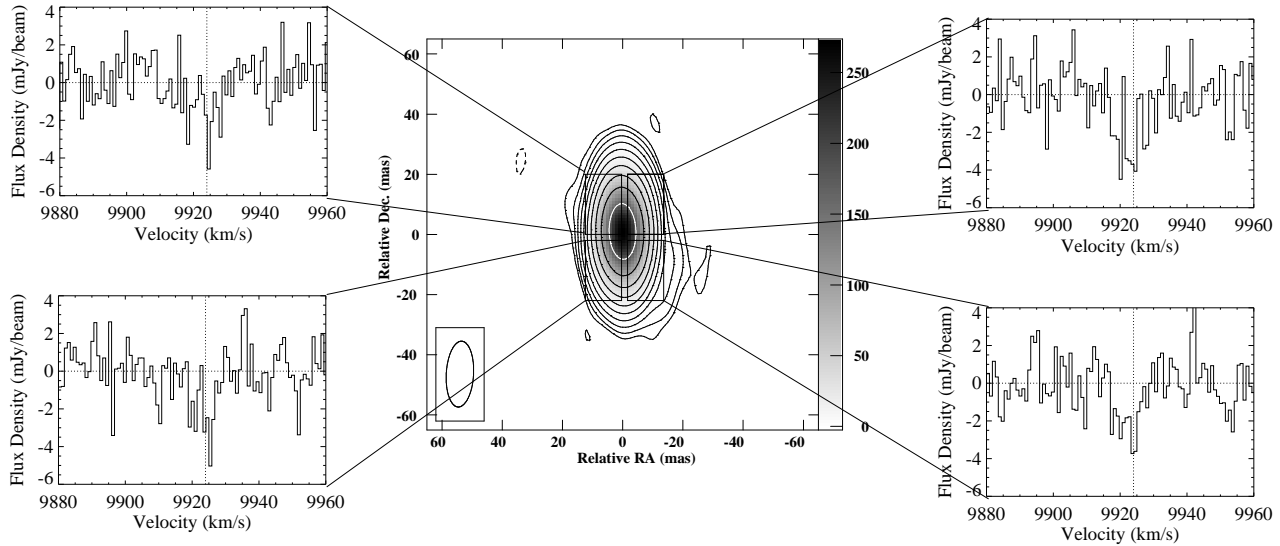}
\caption{{\it Center panel:} The VBLA 21 cm continuum image of the
radio-bright quasar \qsoname\ (grayscale and contours). The continuum
emission contours are plotted in linear increments of -4, 4 8, 16, 32,
64, 128, 256, 512, 1024, and 2048 of the rms flux of 80~$\mu$Jy/beam,
and the grayscale follows the scale bar on the right side of the
central panel; the beam size and orientation are shown in the inset.
The quasar is modestly resolved with peak and integrated flux
densities of 272~($\pm$0.1)~mJy/beam and 341~($\pm$0.2)~mJy,
respectively. {\it Side panels:} \HI 21~cm absorption spectra from the
foreground galaxy \galaxyname\, extracted from the four independent
regions to the background quasar indicated with boxes overlaid on the
continuum image. Each \HI 21 cm absorption panel is connected to the
box indicating the region from which it was extracted. For purposes of
comparison , all of the spectral plots are plotted with the same axis
scales, and the velocity 9924~\kms is marked for reference (the
optical redshift of the foreground galaxy is 9932$\pm$10 km
s$^{-1}$).\label{vlbacontin_spec}}
\end{figure*}

\subsection{Very Long Baseline Array Observations  \label{sec:obs_vlba}}

The VLBA observations of this source were carried out in two 8-hour
observing sessions on the nights of 15 and 16 June 2008 under program
ID BY124.  Four adjacent 4~MHz baseband channel pairs were used in the
observations, with both right and left hand circular polarization, and
sampled at 2 bits.  The data were correlated at the VLBA correlator in
Socorro, New Mexico in two passes.  In the first pass, all the
baseband channels were correlated with 32 spectral channels per 4~MHz
for the purpose of imaging the radio continuum emission.  The second
correlation pass was performed only on the second pair of baseband
channels, which were centered at 9925~km s$^{-1}$.  This produced a data cube with 1024 spectral channels and a resolution of 3.9~kHz (0.9~km s$^{-1}$)
per channel. The total on-source integration time was 15 hours.  Two
of the 10 VLBA antennas were rendered unusable due to technical
problems (Brewster) and data corruption (Mauna Kea). The data
reduction was performed using AIPS.

After a priori flagging of data affected by interference in both data
sets, amplitude calibration was performed using the measurements of
the antenna gains and the system temperatures for each station.
Bandpass calibration was performed using 3C 273.  The continuum data
set was then self-calibrated and imaged in an iterative cycle.  Figure
\ref{vlbacontin_spec} shows the continuum image of the background
quasar \qsoname\ that was obtained using an intermediate grid weighting
between pure natural and pure uniform (Robust=2) in the IMAGR routine
of AIPS.  The angular resolution of this image is 21.9~mas
$\times$~9.1~mas (P.A. = 3$^{\circ}$) and the rms noise is
80~$\mu$Jy/beam.  At this resolution the source is resolved with peak
and integrated flux densities of 272.0~($\pm$0.1)~mJy/beam and
341.0~($\pm$0.2)~mJy, respectively.  The spatial extent of the
continuum source was fitted with a Gaussian profile that gave a
nominal deconvolved size of 8.5~mas~$\times$~2.4~mas at FWHM.  The
radio emission associated with the QSO is core-dominated with an
extension toward the southwest side of the core indicative of a weak
jet-like feature. The visibility amplitude decreases to 50\% of the peak at baseline length of $\sim$10 mega lambda, thus confirming the spatial extension of the QSO.
The integrated flux measured in our VLBA image was
40.6~mJy less than that of the FIRST VLA survey, which suggests the
presence of faint extended structure that was detected by the VLA but
was resolved out by the VLBA.

The self-calibration solutions of the continuum data set were applied
to the spectral line data cube, which was then imaged using a similar
weighting as the continuum. The continuum emission was then subtracted
from the \HI\ cube. The bottom panel of Figure \ref{j1042_gbt_vla_vlba}
shows the un-smoothed VLBA 21~cm \HI\ spectrum from pixels with at least 400~$\mu$Jy (5~$\sigma$) continuum flux. The rms noise in the VLBA spectrum is  2~mJy/beam. Again,
absorption near the redshift of \galaxyname\ is readily apparent.
A comparison of the VLBA and the GBT spectra binned at the same resolution is shown in the lower panel of Figure~\ref{j1042_gbt_zoom}. To estimate the column density, we included channels within the velocity range 9913 to 9930~\kms . In order to minimize the effect of noise in the optical depth cube and the column density measurement, we blanked (removed) the pixels where the continuum emission was less than 100~mJy.

\section{21 cm Measurements\label{sec:measurements}}
 
\subsection{21 cm Component Structure and Kinematics\label{sec:components}}

As shown in Figure~\ref{j1042_gbt_vla_vlba}, the \HI 21~cm absorber in
the foreground of \qsoname\ is confirmed by our observations using the
GBT, the VLA, and the VLBA. While the GBT spectra provide the highest
spectral resolution, the VLBA imaging complements the GBT data by
providing high angular resolution.  Independent observations with the
three telescopes are also helpful for overcoming systematic problems
specific to one facility, e.g., the RFI problem at $v \approx 9880$ km
s$^{-1}$ in the GBT data.

The \HI absorber detected by the GBT has a primary component
consistent with a single Gaussian profile with a centroid at
9923.97~$\pm$~0.17~\kms or 1374.8923~$\pm$~0.0007~MHz ($z_{\rm abs}$ = 0.033103) and a FWHM of
3.6~$\pm$~0.4~km s$^{-1}$.  A weaker component ($\sim 2.3~\sigma$) can
be seen at 9917.6~km s$^{-1}$. 
Close inspection of
Figure~\ref{j1042_gbt_zoom} reveals a possible third component at
$\approx 9927$ km s$^{-1}$, but the significance of this third feature
is marginal.  Figure~\ref{j1042_gbt_zoom} shows the fit of a
single Gaussian profile to the primary component of the \ion{H}{1}
absorption in the continuum subtracted GBT spectrum.  The width of the
Gaussian could be slightly overestimated due to blending with the
tentative third component at 9927 km s$^{-1}$.  Given the marginal
significance of the third component, we have elected to fit the main
component as shown in Figure~\ref{j1042_gbt_zoom} so that our line
width places a conservative upper limit on the kinetic temperature of
the absorbing gas (see \S \ref{sec:dis_phycond}). The bottom panel of Figure~\ref{j1042_gbt_zoom} 
shows a comparison of the GBT and the VLBA spectra. The profiles show similar peak absorption flux and line shapes for the primary component. However, the second component is much more prominent (4.6~$\sigma$) in the VLBA spectrum and is slight shifted in velocity (v=9920~\kms).

The redshift measured from our highest resolution GBT \HI data is
$\sim$8~\kms lower than the redshift derived from the H$\alpha$ data,
but is well within the uncertainty. In
principle, the optical redshift could represent the systemic redshift
of the galaxy.  However, all of our optical spectra (from SDSS and
APO) that detect the foreground galaxy emission lines are measured at
the position of the QSO. As we can see from
Figures~\ref{fig_colorpic1} and \ref{fig_grscale}, this position is
offset from the galaxy center, and it is likely that at this location,
the \ion{H}{2} regions that produce the optical emission lines are
rotating with the disk of the galaxy.  The good agreement of the
optical and 21cm redshifts suggests that the 21cm absorption arises in
the disk and is corotating with the \ion{H}{2} regions and the disk.
The kinematical quiescence of the 21cm-absorbing gas warrants comment
-- as discussed in \S~1, high$-z$ DLAs and sub-DLAs \citep[e.g.,][]{meiring09} are
noted for complex gas kinematics, which likely plays an important role
in how the DLAs evolve.  The 21cm absorption traces the {\it neutral}
gas, and we see no evidence of complex kinematics in the neutral gas
along our sight line through \galaxyname.  We are likely detecting an
ordinary gas cloud in the disk of the galaxy.

\subsection{Spatial Variability of the 21 cm Absorption\label{sec:spatial}}
\begin{figure}
\centerline{\includegraphics[ angle=0,scale=0.7]{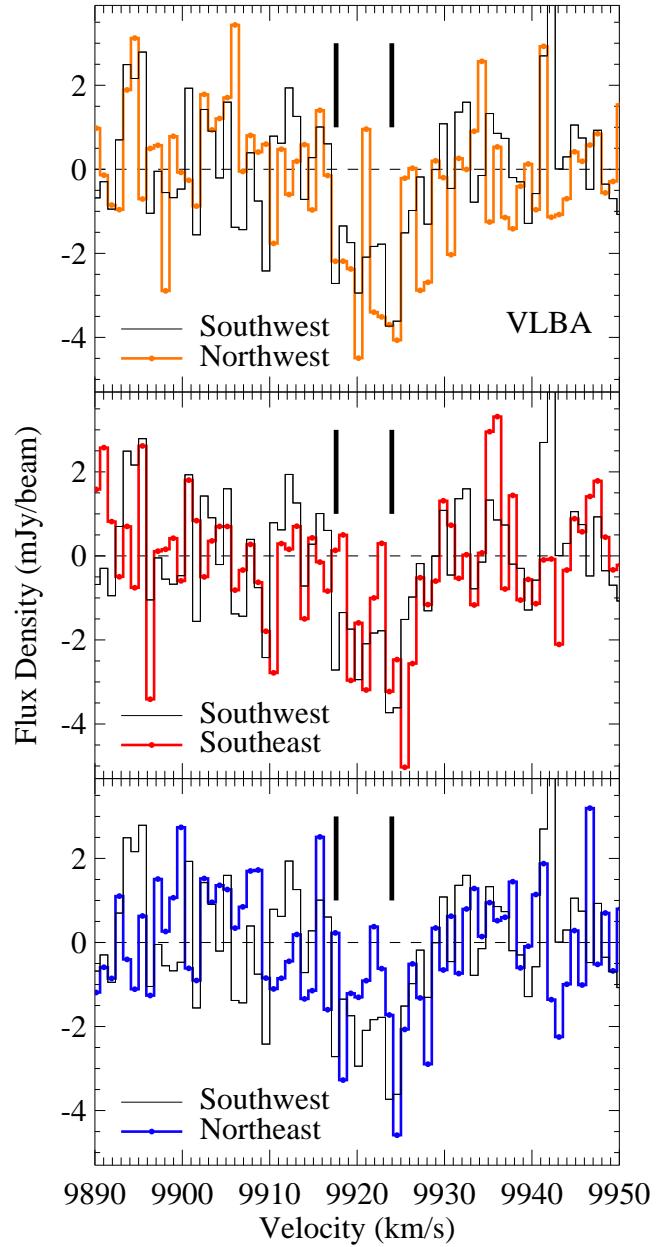}}
\caption{Direct comparison of the continuum-subtracted VLBA 21cm
absorption profiles extracted from the four regions shown in
Figure~\ref{vlbacontin_spec}. {\it Top:} Absorption profile from the
Southwest region (thin black line) overplotted on the profile from the
Northwest region (thick orange line). {\it Middle:} Absorption profile
from the Southwest region (thin black line) overplotted on the profile
from the Southeast region (thick green line). {\it Bottom:} Absorption
profile from the Southwest region (thin black line) overplotted on the
profile from the Northwest region (thick blue line). In each panel,
the thick black vertical lines above the spectra indicate the velocity
centroids of the two components detected with the
GBT. \label{vlba_compare}}
\end{figure}

\begin{figure*}
\includegraphics[angle=-90,scale=0.8]{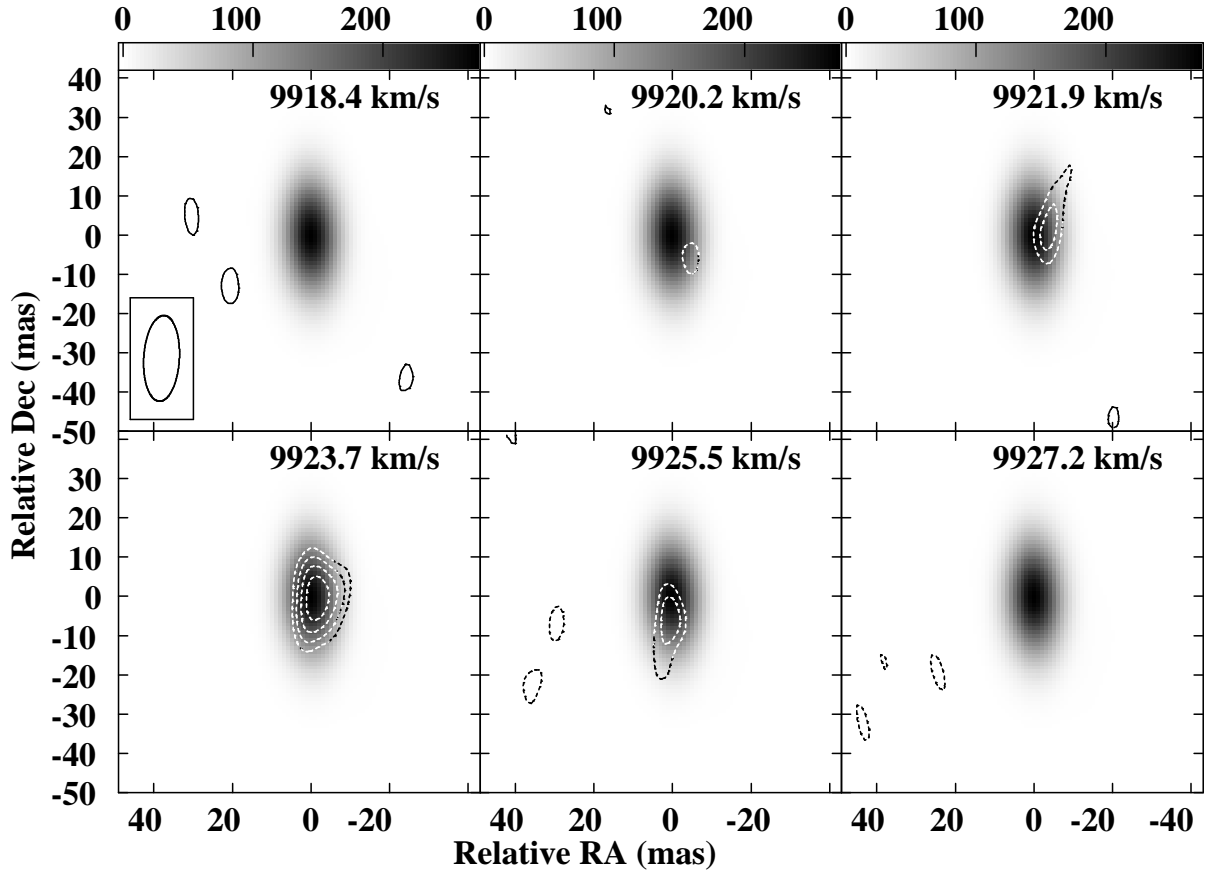}
\caption{The VLBA 21 cm \ion{H}{1} absorption from \galaxyname\
(contours) superimposed on the VLBA 21cm continuum image (greyscale
with the scalebars at top) of the background quasar \qsoname\ in
velocity channel maps extending over the velocity range of the
absorption detected in the GBT spectrum. The central velocity is
indicated in each panel, and the beam characteristics are shown in the
inset at upper left.  To enhance the signal, two VLBA channels have
been combined in each panel. The contours are plotted in linear
increments of 3, 4, 5 and 6 times the rms noise of 1.5~mJy/beam.}
\label{vlbachanmap}
\end{figure*} 

One of the broad goals of this project is to investigate the
tranvserse spatial distribution of the absorbing gas by taking
advantage of the extended nature of QSOs at radio frequencies. For this
purpose, Figures~\ref{vlbacontin_spec}, \ref{vlba_compare}, and
\ref{vlbachanmap} provide more detailed presentations of the spatial
structure of the 21~cm absorption. 
Figure~\ref{vlbacontin_spec} shows
the high resolution VLBA continuum image of the background QSO along
with \HI absorption spectra extracted from four different regions $-$ Northwest, Northeast, Southwest, and
Southeast. Each region is roughly the size of the synthesized beams. 
 To facilitate detailed comparisons, Figure~\ref{vlba_compare}
overplots the spectra from the Northwest, Northeast, and Southeast
regions on top of the spectrum from the Southwest region. From these
comparisons, we notice the following trends: First, as the regions 
move from West to East, the peak absorption flux in the main component
increases and the velocity centroids shift to slightly higher
velocities. Second, the weaker component (at $v
\approx 9920$ km s$^{-1}$) is evident in all of the VLBA
regions with is maximum strength in the West (3.1~$\sigma$ in the Southeast and Southwest) and  weakest in the Northeast (1.8~$\sigma$). Both components appear to be narrower on the Eastern side,
but this could be simply due to more blending on the Western side if
the velocity centroids of the two components are shifting closer
together in the West. The spatial variation in the strength of the
absorber for each spectral channel is shown in
Figure~\ref{vlbachanmap} where the absorption feature contours are
superposed on the continuum image shown in grey scale. Each velocity
plane is independent and the noise should not correlate between the
images.

 In the main component, the absorber covers most of
the background quasar at the resolution of the VLBA imaging. 
The extent of the HI cloud as measured from the region with at least
4~$\sigma$ HI detection is 41~mas~$\times$~21~mas on the plane of the
sky, which corresponds to a physical size of 27.1~pc~$\times$~13.9~pc
at the redshift of the absorber. The similarity in peak absorption flux between the GBT and VLBA profiles suggest that the absorption is associated with the same absorber. Since the GBT beam is much larger, it is possible that the GBT spectrum includes absorption from an entirely different cloud than the gas that imprints the absorption on the VLBA spectrum.  However, we would intuitively expect a separate and independent cloud to have different kinematics, which in turn would cause the GBT absorption profile to have a different shape from the VLBA absorption. Also, in this case, since the QSO is not very extended (the VLBA flux density is just $\approx$40~mJy lower than the flux from VLA FIRST Survey) and the absorber is very narrow, it is unlikely that there are multiple clouds associated with the absorber.

\subsection{Kinetic Temperature of the \HI\ 21~cm Absorber \label{sec:dis_phycond} }

Interstellar neutral gas in disk galaxies is believed to be
distributed in two distinct physical phases - the warm neutral medium
(WNM) and the cold neutral medium (CNM). While the WNM is diffuse and
has a temperature in the range of 5000 to 8000~K, the CNM exists as
clumpy dense clouds with temperatures in the range of 20 to 250~K
\citep{kul_heiles88}. This has been confirmed in numerous
extragalactic studies including \citet{young_lo97a,young_lo97b},
\citet{carilli98}, and \citet{lane00}. 

The width of an absorption line can be used to derive an upper limit
on the kinetic temperature of the gas, assuming the line width is
predominantly due to thermal broadening (this is an upper limit
because other factors, e.g., turbulence, can also broaden the
line). For neutral hydrogen, the kinetic temperature upper limit can
be estimated using
 \begin{equation}{\label{eq-kin_temp}}
 T_{k} \le 21.855 (\Delta v)^2 
 \end{equation}
where $\Delta v$ is the FWHM velocity in \kms. Based on the Gaussian
fitting from our high resolution GBT data, we estimate that the FWHM
of the absorption feature is 3.6~\kms, which indicates that 
\begin{equation}
T_{k} \le 283 \ {\rm K.}
\end{equation}
 Thus, the temperature suggests that the 21cm absorption arises in
the CNM of \galaxyname .

\subsection{H~I Column Density and Volume Density  \label{sec:nhi}}

With the temperature constraint afforded by the high spectral
resolution of the data, the \ion{H}{1} column density can be estimated
by integrating over the 21 cm absorption profile using the standard
equation \citep[e.g.,][]{rohlfs86}:
\begin{equation}
N({\rm H~I}) = 1.823 \times 10^{18} \frac{T_{s}}{f}\int \tau _{21}(v) dv,
\end{equation}
where $T_{s}$ is the spin temperature, $f$ is the fraction of the
radio flux source that is covered by the absorbing gas and is assumed to be unity, and $\tau_{21}
(v)$ is the 21 cm optical depth in velocity space (in km
s$^{-1}$). In the GBT spectra the peak optical depth was measured to be 0.076~$\pm$~0.011. Given
the context of the absorption (i.e., arising in the disk of a spiral
galaxy), it is likely that the density is high enough so that the
level popultions of the \ion{H}{1} hyperfine structure levels are set
predominantly by collisions.  Thus we can assume that $T_{s} \approx
T_{k}$.  With $T_{k} \le$ 283 K (\S \ref{sec:dis_phycond}), we obtain
\begin{equation}
N({\rm H \ I})_{\rm GBT} \le 9.6 \times 10^{19} \ {\rm cm}^{-2} 
\end{equation}
by integrating over the main absorption line detected with the GBT.  This
indicates that the 21cm absorber is not quite a damped \Lya absorber according to the usual definition. This is somewhat
surprising given the small impact parameter of the sight line to an
actively star-forming spiral galaxy. 
We discuss this further in \S~6. However, if we apply the same procedure to the VLBA data to extract the overall \HI\ profile for pixels with continuum flux $\ge$ 100~mJy, we obtain
\begin{equation}
N({\rm H \ I})_{\rm VLBA} \le 1.5 \times 10^{20} \ {\rm cm}^{-2}.
\end{equation}
The difference between the GBT and the VLBA column density constraints
could result from a difference in the covering fraction ($f$) between
the two observations. Since the VLBA resolves out the source, the
angular extent of the continuum source could be smaller for the VLBA
than the GBT, so although the covering fraction estimated from the
VLBA observations is $\sim$1, the covering fraction for the GBT
observations could be lower than 1.  We note from
Figure~\ref{vlba_compare} that while the VLBA centroids of the main
absorption component (at $v \approx ~$9924~\kms\ ) match up well
with the centroid of that component indicated by the GBT data, the GBT
centroid of the weaker feature (at 9917.6 \kms\ ) is slightly
offset to shorter velocities compared to the velocity of that
component in the VLBA spectra.  This could also be an indication that
some of the GBT absorption at 9917.6 \kms\ occurs outside the continuum region detected by the VLBA.

In \S \ref{sec:spatial}, we showed that there is some subtle spatial
variability of the 21cm absorption, but overall, similar absorption is
detected in all directions probed with the VLBA data.  This indicates
that the dimensions of the absorbing cloud are at least 27.1~pc~$\times$~13.9~pc. Assuming the cloud to be an ellipsoid that has a spin temperature of 283~K, the estimated \HI\ mass limit for this cloud is 356~$M_{\odot}$. If we take $\approx 14$ pc as the minimum line-of-sight size of the cloud and combine this with the VLBA upper limit on $N$(\ion{H}{1}), we find that the volume density of the gas is
\begin{equation}
n({\rm H~I}) < 3.5 \ {\rm cm}^{-3}.
\end{equation}
While the cloud size implied by the VLBA data is consistent with
expectations for a CNM cloud, this upper limit on $n$(\ion{H}{1}) is
an order of magnitude lower than expected for the CNM
\citep[cf.,][]{jorgenson09,tielens}. It is worth noting that our density values are based on the assumption that the line-of-sight length of the absorbing cloud is 14~pc. 
Since the VLBA  absorption region of 27~pc$\times$14~pc transverse to the
sight line covers almost the entire continuum source with an extension
in the Northwest-Southeast direction, it may be possible that the
absorbing cloud extends beyond the region probed by the VLBA continuum
data. Therefore, our size measurement is a limiting value and the line-of-sight length of 14~pc is a conservative lower limit.  If the absorbing cloud is larger, then the implied number density is even lower, which further exacerbates the discrepancy with the typical density expected in CNM regions.  Of course, it is possible that the cloud contains high-density internal clumps that are much smaller than the VLBA beam.  In this case, the line-of-sight length could be lower and the density could be higher.  However, in this situation we might expect to see more dramatic spatial variability in the VLBA absorption profiles.  We do notice some weak variability in the VLBA profiles, but overall the four VLBA sight lines  show the same basic profile shapes (see Figure~\ref{vlbacontin_spec}). This suggests that the absorbing cloud extends across the entire region probed by the VLBA data.  
We have not estimated the effects of ISM scintillation and microlensing on the apparent size of the emitting region. Such assessment is beyond the scope of this discussion.

\subsection{Total H~I Mass of \galaxyname \label{sec:total_h1_mass} }

Our 21~cm spectra show no indication of \HI\ emission
associated with the foreground galaxy (see
Figure~\ref{j1042_gbt_vla_vlba}). The absence of 21~cm emission
provides an upper limit on the total \ion{H}{1} mass of \galaxyname .  Based on the GBT noise properties and assuming a line-width of 100~kms$^{-1}$, we estimate a 3$\sigma$ \HI mass limit
of less than 5.1~$\times$~10$^8$ M$_\odot$ for the foreground
galaxy. This is an order of magnitude less that the \HI mass associated with a typical luminous spiral galaxy ($L\sim L_{\star}$). However, \galaxyname\ has $L \ll L_{\star}$, and this
upper limit on $M$(\ion{H}{1}) is entirely consistent with the
\ion{H}{1} masses measured in dwarf galaxies
\citep[e.g.,][]{matthews95,matthews96,matthews98,swaters02,begum08}.
 Our GBT data should provide the most stringent limits on $M$(\ion{H}{1}) for any narrow features with $\Delta V\sim$100~\kms. The GBT spectra suffered from sinusoidal standing waves as commonly found when observing bright continuum sources and consequently broad and faint features of the order  $\Delta V\ge$~300~\kms may have been missed. We also examined the ALFALFA data (private communication with R. Giovannelli) which show signs of a moderate enhancement from 9775 to 10078~\kms . This corresponds to an \HI\ mass of 4.5~$\times$~10$^9~ M_\odot$. However, the ALFALFA data also shows signs of sinusoidal modulations thus adding a significantly large uncertainty to the mass estimation.  We plan to observe \galaxyname\ with the Arecibo Telescope to correctly estimate its \HI mass. The next section discusses the implications of the \HI\  mass of \galaxyname\ and compares the galaxy to \HI\ emission from other dwarf galaxies.

\section{Discussion\label{sec:discussion}}

\subsection{H~I Deficiency: Evidence of Gas Consumption in an Isolated Environment?  \label{sec:gas_comsumption} }

 High-resolution \HI  surveys of dwarf and irregular galaxies such as the Westerbork observations of neutral Hydrogen in Irregular and SPiral galaxies (WHISP) Survey \citep{swaters02} or the Faint Irregular Galaxy GMRT Survey (FIGGS) \citep{begum08} generally find dwarf galaxies to have \HI\ envelopes extending over much larger area than their optical disks.  However, our radio observations of the QSO sight line that pierce the inner disk of \galaxyname\ reveal that the galaxy is
surprisingly \ion{H}{1}-deficient in several ways. Although we detect a cold \ion{H}{1} cloud in this galaxy, it is a  sub-DLA at best, unlike what is expected from emission maps of similar galaxies from the WHISP and FIGGS  surveys. Likewise, by combining the \ion{H}{1} column density with constraints on the size of the absorbing cloud from the VLBA data, we find that $n$(\ion{H}{1}) $<$ 3.5 cm$^{-3}$, which is roughly an order of magnitude lower than expected for a CNM cloud. Here, we attempt to understand if \galaxyname\ , which appears to be an ordinary dwarf spiral in optical imaging with ongoing star formation, is consistent with \HI\ emission studies of dwarf galaxies or if \galaxyname~ is a special case.   

 In order to understand the discrepancy between dwarf \HI\ galaxies and \galaxyname,~ it is important to analyze any possible bias between our selection criterion and that of the above mentioned surveys. Since \galaxyname~ was identified using optical emission lines, our selection criteria differ from those of WHISP and FIGGS, which are {\it \HI$-$selected}, i.e., a criterion for including the galaxies in these surveys in the first place was that they were already known to have detectable \ion{H}{1} emission. This could introduce a problematic bias when comparing their properties to \galaxyname. It is possible that galaxies such as \galaxyname~ belong to a different population of dwarf galaxies than the \HI\-rich dwarfs commonly seen in \HI\ surveys.

 We believe the most appropriate way to understand the nature of \galaxyname~ is to compare it with a dwarf galaxy sample which was not  \ion{H}{1}$-$selected.  In a series of papers, Matthews and Gallagher surveyed the \ion{H}{1} emission properties of a sample of faint ``extreme late-type'' galaxies selected to have no published \ion{H}{1} information before their survey \citep[e.g.,][]{gallagher95,matthews95,matthews96}. They found a range of modestly to highly gas-rich galaxies with \ion{H}{1} masses and optical properties consistent with our constraints on \galaxyname. In a follow-up study of {\it isolated} extreme late-type galaxies, \citet{matthews98} argue that these galaxies are not ``scaled-down'' versions of luminous gas-rich galaxies.  Instead, they find that the extreme late-type galaxies often form stars sluggishly compared to more luminous late-type galaxies.

These results of \citet{matthews98} suggest a possible class of dwarf galaxies with properties similar to \galaxyname. In \S \ref{sec:galenv}, we showed that \galaxyname\ is an isolated galaxy at the edge of a large-scale structure (see Figure~\ref{sdssslice}). If this dwarf galaxy is forming stars relatively slowly like the extreme late-type galaxies of \citet{matthews98}, it may be gradually consuming its gas reservoir without being replenished with fresh gas from its intergalactic surroundings or from interactions with other galaxies. 

A variation of this hypothesis is that for some reason, this galaxy is not able to efficiently transport gas into the inner region probed by the \qsoname\ sight line so that the \ion{H}{1} is depleted in its central regions thereby creating an \HI\ gap/hole. Interestingly, similar \HI\ gaps/holes have been reported in   \HI$-$selected dwarf galaxies such as DDO43  by \citet{begum08}.  In addition, close inspection of the WHISP \ion{H}{1} maps reveals other galaxies with similar patchy \ion{H}{1} distributions with lower \ion{H}{1} intensities in some parts of their inner regions. It would be interesting to re-observe \galaxyname~ galaxy
with higher angular resolution with the upcoming Atacama Large Millimeter Array to explore the distribution and physical properties of the molecular gas.  High-resolution {\it Hubble Space Telescope (HST)} imaging would also provide insight about the star-formation history in \galaxyname.

\subsection{Quiescent Neutral Gas \label{sec:quiescent_gas} }

Another possible explanation for the low amount of \ion{H}{1} revealed
by our observations is that gas in the central region of the galaxy
has been partially evacuated by a galactic outflow, either a bound
galactic fountain or an escaping galactic wind.  Such outflows have
been observed in nearby starburst galaxies \citep[][and references
therein]{veilleux05}, and indications of a {\it bound} outflow have
been seen in the central region of the Milky Way
\citep[e.g.,][]{bc03,keeney06}. It has long been known that there is a
deficit of extraplanar \ion{H}{1} in the inner 3 kpc of the Milky Way
\citep{lockman84}, which could be due to expulsion of \ion{H}{1} from
the inner Galaxy by some type of outflow \citep[e.g.,][]{everett08}.

If this type of outflow is driving gas out of the inner region of
\galaxyname, then we might expect to see kinematical evidence of the
outflowing gas \citep[as was done in, e.g.,][]{keeney06}.  However,
contrary to the kinematics seen in the metal-line absorption profiles
of many high$-z$ DLAs \citep[e.g.,][]{pw97}, the 21~cm \HI line
profile of \galaxyname\ is quite simple, and we see no indications of
outflowing neutral gas.  However, there are some caveats.  We might
not detect the outflowing neutral gas if the spin temperature is too
high or the covering factor is low.  Conversely, the kinematics of
high$-z$ galaxies are often traced by species, such as \ion{Si}{2},
that can exist in ionized gas as well as neutral gas, and it is possible
that the complex kinematics of the high$-z$ systems could be partly
due to ionized gas that we would not see in 21cm absorption.  To test
this possibility, it would be valuable to observe \qsoname\ in the
ultraviolet with the Cosmic Origins Spectrograph (COS) on {\it HST}.
Ultraviolet spectra provide sensitive probes of ionized gas, and the
combination of the UV and 21cm data would provide a direct measurement
of the spin temperature of the neutral gas. It is also plausible that
a warm neutral medium (WNM) is present in this galaxy; if the WNM 
has a sufficiently large velocity dispersion and low optical depth, then such
features could be lost in the process of baseline fitting. This
can also be tested with COS observations using \ion{O}{1} absorption
lines, which are locked to the neutral gas by a resonant charge
exchange reaction.

In \S~1, we discussed the need to probe the signatures of
neutral gas that exists in different contexts.  Of course, this
requires a sample that is large enough to support statistically
significant conclusions, but it is interesting to note that if we
make the reasonable assumption that $T_{k} \approx T_{s}$ and we adopt
the metallicity from the optical emission lines (Table
\ref{tbl-emlines}), then we find that the spin temperature in
\galaxyname\ adheres to the spin temperature -- metallicity relation
recently presented by \citet{kanekar09b}.  They discuss several
hypotheses for the cause of this correlation and conclude that the increase in the number of possible radiative pathways for cooling gas with an increase in metallicity is the likely cause. Extending their argument to the case of 
\galaxyname, it seems likely that the higher metallicity in the inner
disk of \galaxyname\ causes the gas to cool more rapidly.  Many high-z
DLAs are known to have higher spin temperatures \citep{kanekar05}.
The high$-T_{s}$ absorbers could originate in outer disks (or
tidal/dynamical debris) where the densities and metallicities are
lower and the cooling times are longer.  However, it is also possible
that the cooling times are longer simply because high$-z$ galaxies
have substantially lower metallicities.  It would be interesting to
place constraints on $T_{s}$ in a sample of {\it nearby} galaxies
probed at a range of impact parameters and with a range of galaxy
properties.

\section{Summary and Concluding Remarks\label{sec:conclusion}}

The SDSS presents a remarkable (and largely
untapped) opportunity to study gas that is difficult to observe in
low-redshift galaxies/galaxy groups by observing absorption imprinted
on the spectra of quasars that happen to be located in the background
of galaxies of interest.  While the SDSS obtains spectra 
of these background quasars, the spectra have
somewhat low resolution and sensitivity for absorption spectroscopy
and only cover the optical band.  Thus, while the SDSS provides an
invaluable database for selecting background QSO - foreground galaxy
groupings for this type of study, follow-up observations with higher
resolution instruments and in other frequency bands will be required to
fully exploit the technique.  To demonstrate the potential of SDSS 
for 21cm studies of this type, we have observed with the GBT, VLA, and VLBA
 the radio-loud quasar
\qsoname\ that pierces a foreground star-forming spiral galaxy (\galaxyname) at a very small impact parameter.  To
complement the 21cm observations, we have also obtained images with
the SOAR telescope (which provides better angular resolution than the
SDSS imaging) and follow-up spectroscopy with the APO 3.5m telescope.
From this suite of observations, we have obtained the following
results:

\begin{enumerate}
\item The high-resolution optical imaging with SOAR shows that the
foreground galaxy is a low-luminosity spiral galaxy, and the quasar
sight line is at a projected distance of 2.5$^{\prime\prime}$ from the
center of the foreground galaxy, which corresponds to an impact
parameter of 1.7 kpc for our assumed cosmology.  The emission lines of
the foreground galaxy imply a SFR surface density of
0.004~$M_{\odot}$~yr$^{-1}$ kpc$^{-2}$ and a metallicity of
-0.27$\pm$0.05 dex relative to Solar.  The optical H$\alpha$/H$\beta$
ratio also indicates that the galaxy has a low dust content, at least 
in the region encompassed by the 3$^{\prime\prime}$ SDSS fiber (3.1~kpc$^2$) 
centered at the position of the background QSO. The galaxy colors indicate
 that its approximate stellar mass is M$_{\rm stars} \approx 10^{9.1}$
M$_{\odot}$.

\item Using larger-scale information about the distribution of galaxies
   in the vicinity of the foreground galaxy from the SDSS, we find that
   this object is a relatively isolated galaxy near the boundary of a
   large-scale structure.

\item The spectra obtained with the GBT, VLA, and VLBA independently reveal
\ion{H}{1} 21~cm absorption in the spectrum of the background QSO at
the redshift of foreground galaxy.
Two components separated by $\approx 6$ km s$^{-1}$ are evident in the
GBT and VLBA spectra; a third feature may be present but is only
recorded at marginal significance. However, it is clear that the absorption
profiles indicate simple and quiescent kinematics.  The high spectral
resolution of the GBT data places a strong upper limit on the kinetic
temperature of the 21cm-absorbing gas in the strongest component,
$T_{k} \le 283 \ {\rm K.}$

\item The background QSO is relatively compact, but nevertheless the
   VLBA observations resolve the continuum emission source
   and enable a preliminary search for small-scale spatial variability
   in the 21cm absorption arising in the foreground galaxy.  We find
   variations in the optical depth and centroids of the 21cm
   absorption from four independent sight lines through \galaxyname\
   extracted from the VLBA data, but the variations are at a low
   level, and overall the absorption detected in the four regions shows
   similar optical depths and component structure.  This indicates
   that the main absorbing cloud covers most of the continuum region
   detected with the VLBA and has dimensions of at least 27.1 pc $\times$
   13.9 pc.

\item Combining the upper limit on the kinetic temperature with the
large covering factor indicated by the VLBA data, we obtain
$N$(\ion{H}{1}) $< 9.6 \times 10^{19}$ cm$^{-2}$ from the GBT data and
$N$(\ion{H}{1}) $< 1.5 \times 10^{20}$ cm$\ ^{-2}$ from the VLBA data;
the difference may be due to differences in the region probed by the
large-beam single-dish telescope vs. the small-beam
interferometer. The lower limit on the size of the main absorbing
cloud combined with the VLBA $N$(\ion{H}{1}) constraint indicates that
the \ion{H}{1} volume density is less than 3.5 cm$^{-3}$.


\item We offer some remarks about the implications of our
measurements.  We suggest that \galaxyname\ provides information about
how the stars and gas in an isolated galaxy evolve if left largely
undisturbed.  It appears that the gas in the inner region of the
galaxy is being depleted (compared to other spirals) by conversion to
stars without being replenished with inflowing matter.  It seems
unlikely that an outflow is depleting the gas; we see no evidence of
outflowing material, but further observations are required to properly
search for such outflows.  The galaxy follows the spin temperature --
metallicity relation seen in higher-redshift DLAs.

\end{enumerate}

The present understanding of \HI clouds and their characteristics on
parsec scales outside our own galaxy is quite limited. This is
pathfinding work for future 21~cm absorber investigations. With
upcoming facilities like the EVLA providing higher spectral resolution, larger bandwidth, and smaller beam size,  21cm absorbers will likely be
detected in galaxies toward background QSOs much more efficiently. Moreover, RFI environment for EVLA is expected to be different from single-dish instruments, which may be very helpful in cases affected by local RFIs. This will enable detailed studies of
the nature and distribution of cold gas in galaxies at different
distances from the disk. This is also a useful technique for
assembling a DLA/sub-DLA sample for future UV and/or optical
spectroscopic studies that is not metallicity biased. Combining our
understanding of the cold component traced by 21~cm \HI\ absorbers
with the warmer component traced by \Lya and metal absorbers will be
valuable for understanding the nature, physical
conditions, dynamics, and the role of neutral clouds in
galaxy evolution.

\acknowledgements This research has benefited from discussions with D. Calzetti, J. Gallagher, R. Giovannelli, D. Keres, S. Stanimirovi\'{c}, J. Stocke, and J. van Gorkom. The authors are
grateful to the observatory staff at the GBT, the VLA and the VLBA who
made these observations possible.  SB is grateful for a Student
Observing Support award (GSSP08-0024) from the National Radio Astronomy Observatory
that made this work possible. SB and TMT also acknowledge financial
support for this research from NASA grant NNX08AJ44G. D.V.B is funded
through NASA LTSA grant NNG05GE26G. Funding for the SDSS and SDSS-II
has been provided by the Alfred P. Sloan Foundation, the Participating
Institutions, the National Science Foundation, the U.S. Department of
Energy, the National Aeronautics and Space Administration, the
Japanese Monbukagakusho, the Max Planck Society, and the Higher
Education Funding Council for England. The SDSS Web Site is
http://www.sdss.org/.  The SDSS is managed by the Astrophysical
Research Consortium for the Participating Institutions. The
Participating Institutions are the American Museum of Natural History,
Astrophysical Institute Potsdam, University of Basel, University of
Cambridge, Case Western Reserve University, University of Chicago,
Drexel University, Fermilab, the Institute for Advanced Study, the
Japan Participation Group, Johns Hopkins University, the Joint
Institute for Nuclear Astrophysics, the Kavli Institute for Particle
Astrophysics and Cosmology, the Korean Scientist Group, the Chinese
Academy of Sciences (LAMOST), Los Alamos National Laboratory, the
Max-Planck-Institute for Astronomy (MPIA), the Max-Planck-Institute
for Astrophysics (MPA), New Mexico State University, Ohio State
University, University of Pittsburgh, University of Portsmouth,
Princeton University, the United States Naval Observatory, and the
University of Washington.  The SOAR Telescope is a joint project of:
Conselho Nacional de Pesquisas CientÃ­ficas e
Tecnol\'{o}gicas CNPq-Brazil, The University of North Carolina at
Chapel Hill, Michigan State University, and the National Optical
Astronomy Observatory.

{\it Facilities:} \facility{ARC ()}, \facility{GBT ()}, \facility{Sloan ()}, \facility{VLA ()}, \facility{VLBA ()}

\end{document}